\documentclass[graybox]{svmult}

\usepackage{natbib}
\usepackage{mathptmx}       
\usepackage{helvet}         
\usepackage{courier}        
\usepackage{type1cm}        
\usepackage{makeidx}         
\usepackage{graphicx}        
\usepackage{multicol}        
\usepackage[bottom]{footmisc}
\newcommand{\be}{\begin{equation}}
\newcommand{\ee}{\end{equation}}
\newcommand{\bea}{\begin{eqnarray}}
\newcommand{\eea}{\end{eqnarray}}

\makeindex             

\begin{document}

\title*{Cosmic Ray transport in turbulent magnetic field}
\author{Huirong Yan}
\institute{Huirong Yan \at KIAA, Peking University, Beijing 100871, China, \email{hryan@pku.edu.cn}}
\maketitle

\abstract{Cosmic ray (CR) transport and acceleration are determined by the properties of magnetic turbulence. Recent advances in MHD turbulence call for revisions in the paradigm of cosmic ray transport. We use the models of magnetohydrodynamic turbulence that were tested in numerical simulation, in which turbulence is injected at large scale and cascades to small scales. We shall address the issue of the transport of CRs, both parallel and perpendicular to the magnetic field.
We shall demonstrate compressible fast modes are the dominant cosmic ray scatterer from both quasilinear and nonlinear theories. We shall also show that the self-generated wave growth by CRs is constrained by preexisting turbulence and discuss the process in detail in the context of shock acceleration at supernova remnants and their implications.  In addition, we shall dwell on the nonlinear growth of kinetic gyroresonance instability of cosmic rays induced by large scale compressible turbulence. The gyroresonance of cosmic rays on turbulence is demonstrated an important scattering mechanism in addition to direct interaction with the compressible turbulence. The feedback of the instability on large scale turbulence should be included in future simulations.}

\section{Introduction}
The propagation and acceleration
of cosmic rays (CRs) are governed by their interactions
with magnetic fields. Astrophysical magnetic fields are turbulent and, 
therefore, the resonant and non-resonant (e.g. transient time damping, or TTD)
interaction of cosmic rays with MHD turbulence is the accepted
 principal mechanism to scatter and isotropize
cosmic rays \citep[see][]{Schlickeiser02}. In addition, efficient scattering is essential for the acceleration of cosmic rays. 
For instance, scattering of cosmic rays back into the shock is a
vital component of the first order Fermi acceleration \citep[see][]{Longairbook}. At the same time, stochastic acceleration by turbulence is 
entirely based on scattering. The dynamics of cosmic rays in MHD turbulence holds the key to all high energy astrophysics and related problems. 

We live in an exciting era when we are starting to test fundamental processes taking place at the beginning of the Universe, at the event horizon of black holes, when the nature of dark matter and dark energy is being probed etc. Using computers many researchers make sophisticated complex models to confront the observations in unprecedented details. In the mean time, with the launching of the new facilities, we have much more observational data available than ever before. For instance, CHANDRA observations of supernova
remnants provide a strong constraint to diffusion coefficients and/or magnetic fields near the shock \citep[see, e.g.][]{Bamba05, PYL05}; 
the diffuse
gamma-ray measurements from Fermi from the Galactic disc have been successfully used to
phenomenologically constrain numerical modeling of cosmic rays, e.g., with GALPROP \citep{Ackermann12}; observations of solar
energetic particles (SEP) have been also fruitful over the past decades and lead to better understanding of transport in the solar
wind \citep[see a review by][and references therein]{Horbury05_SEP}. These developments make it urgent that we understand the key physical processes underlying astrophysical phenomena, can parameterize them and, if necessary, use as a subgrid input in our computer models.


At present, the propagation of the CRs is an advanced theory, which makes
use both of analytical studies and numerical simulations. However,
these advances have been done within the turbulence paradigm which
is being changed by the current research in the field.
Instead of the empirical 2D+slab model of turbulence, numerical
simulations suggest anisotropic Alfv\'enic modes following \cite[GS95]{GS95} scalings (an analog of 2D, but not an
exact one, as the anisotropy changes with the scale involved) + fast modes \citep{CL02_PRL}. These progresses resulted in important revisions on the theory of cosmic ray transport (see review by \citealt{LBYO} and references therein). The GS95 turbulence injected on large scales and its extensions to compressible medium is less efficient in scattering of CRs compared to the estimates made assuming that magnetic turbulence consists of plane waves moving parallel to magnetic field \citep{Chandran00, YL02}. Fast compressible modes, on the other hand, are demonstrated as the dominant scattering agent in spite of various damping processes they are subjected to \citep{YL02, YL04, YL08}

At the same time, one should not disregard the possibilities of generation of additional perturbations on small scales by CR themselves. For instance,  the slab Alfv\'enic perturbation can be created, e.g., via streaming instability \citep[see][]{Wentzel74, Cesarsky80}. Instabilities induced by anisotropic distribution of CRs were also suggested as a possibility to scatter CRs \citep[]{Lerche, Melrose74}. Particularly at shock front, studies of instabilities have been one of the major efforts since the acceleration efficiency is essentially determined by the confinement at the shock front and magnetic field amplifications. Examples of the new developments in the field include, current driven instability \citep{Bell2004}, vorticity generation at curved shock \citep{Giac_Jok2007}, through Baroclinic effect \citep{Inoue09}, through precursor \citep{BJL09}, etc. This field is rich in its own and we shall not dwell upon it in this chapter.

In fact, the small scale instabilities and large scale turbulence are not independent of each other. {\em First} of all, the instability generated waves can be damped through nonlinear interaction with the large scale turbulence \citep[henceforth YL02, YL04]{YL02, YL04}. In the case of anisotropic GS95 turbulence, the efficiency is reduced \citep{FG04}. Nonetheless, owing to the non-linear damping, the instabilities can only grow in a limited range, e.g., $\sim< 100$GeV in interstellar medium for the streaming instability \citep{FG04, YL04}. 
{\em Secondly}, the large scale compressible turbulence also generate small scale waves through firehose, gyroresonance instability, etc \citep{Schek06, LB06, YL11, Santos-Lima}. 

Propagation of CRs perpendicular to mean magnetic field
is another important problem for which one needs to take into account both large and small scale interactions in tested models of turbulence. Indeed, if one takes only the diffusion along the magnetic field line and field line random walk \citep[FLRW][]{Jokipii1966, Jokipii_Parker1969, Forman1974}, compound (or subdiffusion) would arise. Whether the subdiffusion is realistic in fact depends on the models of turbulence chosen \citep{YL08, Yan:2011valencia}. In this chapter we review current understandings to this question within the domain of numerically tested models of MHD turbulence.

In what follows, we introduce the basic mechanisms for the interactions between particles and turbulence in \S2. We discuss the cosmic ray transport in large scale turbulence, including both analytical and numerical studies in \S3. Applications to cosmic ray propagation is presented in \S4. In \S5, we consider the perpendicular transport of cosmic rays on both large and small scales. We shall also discuss the issue of super-diffusion and the applicability of sub-diffusion. In \S6, we concentrate on the issue of self-confinement in the presence of preexisting turbulence and dwell on, in particular, the streaming instability at supernova remnant shocks and its implication for CR acceleration. \S7, we address the issue of gyroresonance instability of CRs and its feedback on large scale compressible turbulence. Summary is provided in \S8.

\section{Interactions between turbulence and particles}
\label{basics}
Basically there are
two types of resonant interactions: gyroresonance acceleration
and transit acceleration (henceforth TTD). The resonant condition is $\omega-k_{\parallel}v\mu=n\Omega$ ($n=0, \pm1,2...$),
where $\omega$ is the wave frequency, $\Omega=\Omega_{0}/\gamma$
is the gyrofrequency of relativistic particle, $\mu=\cos\xi$,
where $\xi$ is the pitch angle of particles. TTD formally corresponds to $n=0$ and it requires compressible perturbations.  

The Fokker-Planck equation is generally used to describe
the evolvement of the gyrophase-averaged distribution function $f$,

\[
\frac{\partial f}{\partial t}=\frac{\partial}{\partial\mu}\left(D_{\mu\mu}\frac{\partial f}{\partial\mu}+D_{\mu p}\frac{\partial f}{\partial p}\right)+\frac{1}{p^{2}}\frac{\partial}{\partial p}\left[p^{2}\left(D_{\mu p}\frac{\partial f}{\partial\mu}+D_{pp}\frac{\partial f}{\partial p}\right)\right],\]
 where $p$ is the particle momentum. The Fokker-Planck coefficients
$D_{\mu\mu},D_{\mu p},D_{pp}$ are the fundamental physical parameters
for measuring the stochastic interactions, which are determined by
the electromagnetic fluctuations \citep[see][]{SchlickeiserMiller}:

Gyroresonance happens when the Doppler shifted wave frequency matches the Larmor frequency of a particle. In quasi-linear theory (QLT), the Fokker-Planck
coefficients are given by  \citep[see][]{SchlickeiserMiller, YL04}

\begin{eqnarray}
\left(\begin{array}{c}
D_{\mu\mu}\\
D_{pp}\end{array}\right)  =  {\frac{\pi\Omega^{2}(1-\mu^{2})}{2}}\int_{\bf k_{min}}^{\bf k_c}dk^3\delta(k_{\parallel}v_{\parallel}-\omega \pm \Omega)
\left[\begin{array}{c}
\left(1+\frac{\mu V_{ph}}{v\zeta}\right)^{2}\\
m^{2}V_{A}^{2}\end{array}\right]\times\nonumber\\
\times\left\{ \left[J_{2}^{2}\left({\frac{k_{\perp}v_{\perp}}{\Omega}}\right)+J_{0}^{2}\left({\frac{k_{\perp}v_{\perp}}{\Omega}}\right)\right]
\left[\begin{array}{c}
M_{{\mathcal{RR}}}({\mathbf{k}})+M_{{\mathcal{LL}}}({\mathbf{k}})\\
K_{{\mathcal{RR}}}({\mathbf{k}})+K_{{\mathcal{LL}}}({\mathbf{k}})\end{array}\right]\right.\nonumber\\
\left.-2J_{2}\left({\frac{k_{\perp}v_{\perp}}{\Omega}}\right)J_{0}\left({\frac{k_{\perp}v_{\perp}}{\Omega}}\right)
\left[e^{i2\phi}\left[\begin{array}{c}
M_{{\mathcal{RL}}}({\mathbf{k}})\\
K_{{\mathcal{RL}}}({\mathbf{k}})\end{array}\right]+e^{-i2\phi}\left[\begin{array}{c}
M_{{\mathcal{LR}}}({\mathbf{k}})\\
K_{{\mathcal{LR}}}({\mathbf{k}})\end{array}\right]\right]\right\} ,\label{gyro}
\end{eqnarray}
where $\zeta=1$ for Alfv\'{e}n modes and $\zeta=k_{\parallel}/k$
for fast modes, $k_{min}=L^{-1}$, $k_c=\Omega_{0}/v_{th}$
corresponds to the dissipation scale, $m=\gamma m_{H}$ is the relativistic
mass of the proton, $v_{\perp}$ is the particle's velocity component
perpendicular to $\mathbf{B}_{0}$, $\phi=\arctan(k_{y}/k_{x}),$
${\mathcal{L}},{\mathcal{R}}=(x\pm iy)/\sqrt{2}$ represent left and
right hand polarization. $M_{ij}$ and $K_{ij}$ are the correlation tensors of magnetic and velocity fluctuations.  

From the resonance condition, we know that the most important interaction
occurs at $k_{\parallel}=k_{\parallel,res}=\Omega/v_{\parallel}$.
This is generally true except for small $\mu$ (or scattering near
$90^{\circ}$). 

TTD happens due to the resonant interaction with parallel magnetic mirror force. Particles can be accelerated by when they are in phase with the waves either by interacting with oscillating parallel electric field (Landau damping), or by moving magnetic mirrors (TTD). When particles are trapped by moving in the same speed with waves, an appreciable amount of interactions can occur between waves and particles. Since head-on collisions are more frequent than that trailing collisions, particles gain energies. Different from gyroresonance, the resonance function of TTD is broadened even for CRs with small pitch angles. The formal resonance peak $k_{\parallel}/k=V_{ph}/v_{\parallel}$ favors quasi-perpendicular modes. However, these quasi-perpendicular modes cannot form an effective mirror to confine CRs because the gradient of magnetic perturbations along the mean field direction $\nabla_{\parallel}\mathbf{B}$ is small. As we will show later in \S\ref{NLT_sec}, the resonance is broadened in nonlinear theory \citep[see][]{YL08}.

\section{Scattering of cosmic rays}
\label{scattering}

\subsection{Scattering by Alfv\'{e}nic turbulence}
\label{Alf_scatter}
As we discussed in $\S$2, Alfv\'{e}n modes are anisotropic, eddies
are elongated along the magnetic field, i.e., $k_{\perp}>k_{\parallel}$.
The scattering of CRs by Alfv\'{e}n modes is suppressed first because
most turbulent energy goes to $k_{\perp}$ due to the anisotropy of
the Alfv\'{e}nic turbulence so that there is much less energy left
in the resonance point $k_{\parallel,res}=\Omega/v_{\parallel}\sim r_{L}^{-1}$.
Furthermore, $k_{\perp}\gg k_{\parallel}$ means $k_{\perp}\gg r_{L}^{-1}$
so that cosmic ray particles have to be interacting with lots of eddies
in one gyro period. This random walk substantially decreases the scattering
efficiency. The scattering by Alfv\'en modes was studied in YL02. In case that the pitch angle $\xi$ not close to 0, the analytical result is  \begin{equation}
\left[\begin{array}{c}
D_{\mu\mu}\\
D_{pp}\end{array}\right]=\frac{v^{2.5}\mu^{5.5}}{\Omega^{1.5}L^{2.5}(1-\mu^2)^0.5}\Gamma[6.5,k_c^{-\frac{2}{3}}k_{\parallel,res}L^{\frac{1}{3}}]\left[\begin{array}{c}
1\\
m^{2}V_{A}^{2}\end{array}\right],\label{ana}\end{equation}
where $\Gamma[a,z]$ is the incomplete gamma function. The presence
of this gamma function in our solution makes our results orders of
magnitude larger than those%
\footnote{The comparison was done with the resonant term in Chandran (2000) as the nonresonant term is spurious %
} in \cite{Chandran00}, who employed
GS95 ideas of anisotropy, but lacked the quantitative
description of the eddies. However,
the scattering frequency,

\be
\nu=2D_{\mu\mu}/(1-\mu^{2}),\label{nu}
\label{nu}
\ee
are nearly $10^{10}$ times lower than the estimates for isotropic and slab model (see Fig.~\ref{impl} {\em left}). {\em It is clear that for most interstellar circumstances, the scattering by Alfv\'enic turbulence is suppressed.} As the anisotropy of the Alfv\'{e}n modes is increasing with the
decrease of scales, the interaction with Alfv\'{e}n modes becomes
more efficient for higher energy cosmic rays. When the Larmor radius
of the particle becomes comparable to the injection scale, which is
likely to be true in the shock region as well as for very high energy cosmic
rays in diffuse ISM, Alfv\'{e}n modes get important.

\subsection{Cosmic ray scattering by compressible MHD turbulence}

As we mentioned earlier, numerical simulations of MHD turbulence supported the GS95 model of turbulence,
which does not have the "slab" Alfv\'enic modes that produced most of the scattering in the earlier models
of CR propagation. Can the turbulence that does not appeal to CRs back-reaction (see \S 4) produce 
efficient scattering? 

In the models of ISM turbulence \citep[]{Armstrong95, Mckee_Ostriker2007}, where the injection happens at large scale, 
fast modes were identified as a scattering agent for cosmic rays in interstellar medium \cite[]{YL02,YL04}.
These works made use of the quantitative description of turbulence
obtained in \cite{CL02_PRL}  to calculate
the scattering rate of cosmic rays. 

\begin{figure*} [h!t] 
{\includegraphics[width=0.45\textwidth]{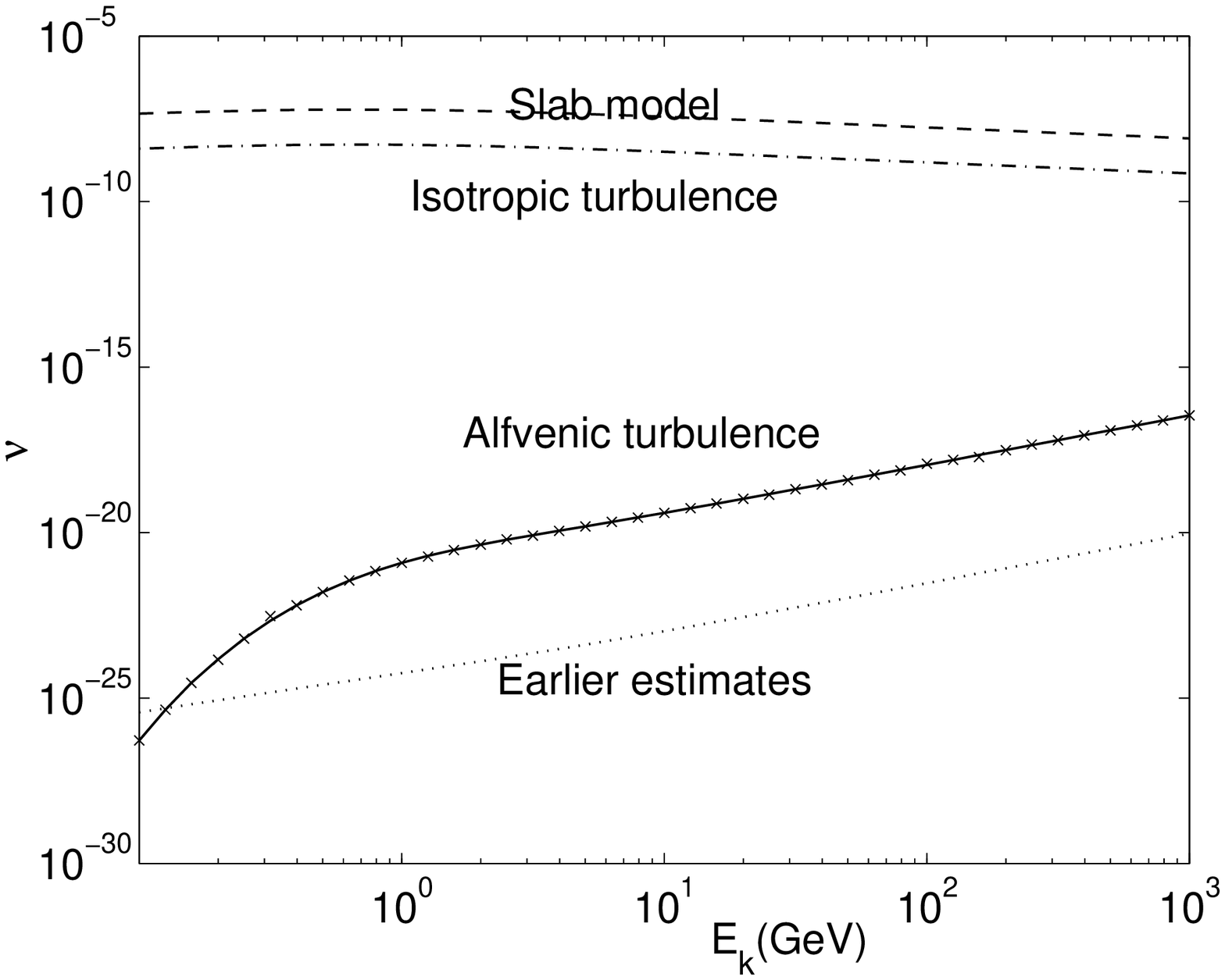} 
\includegraphics[width=0.45\textwidth]{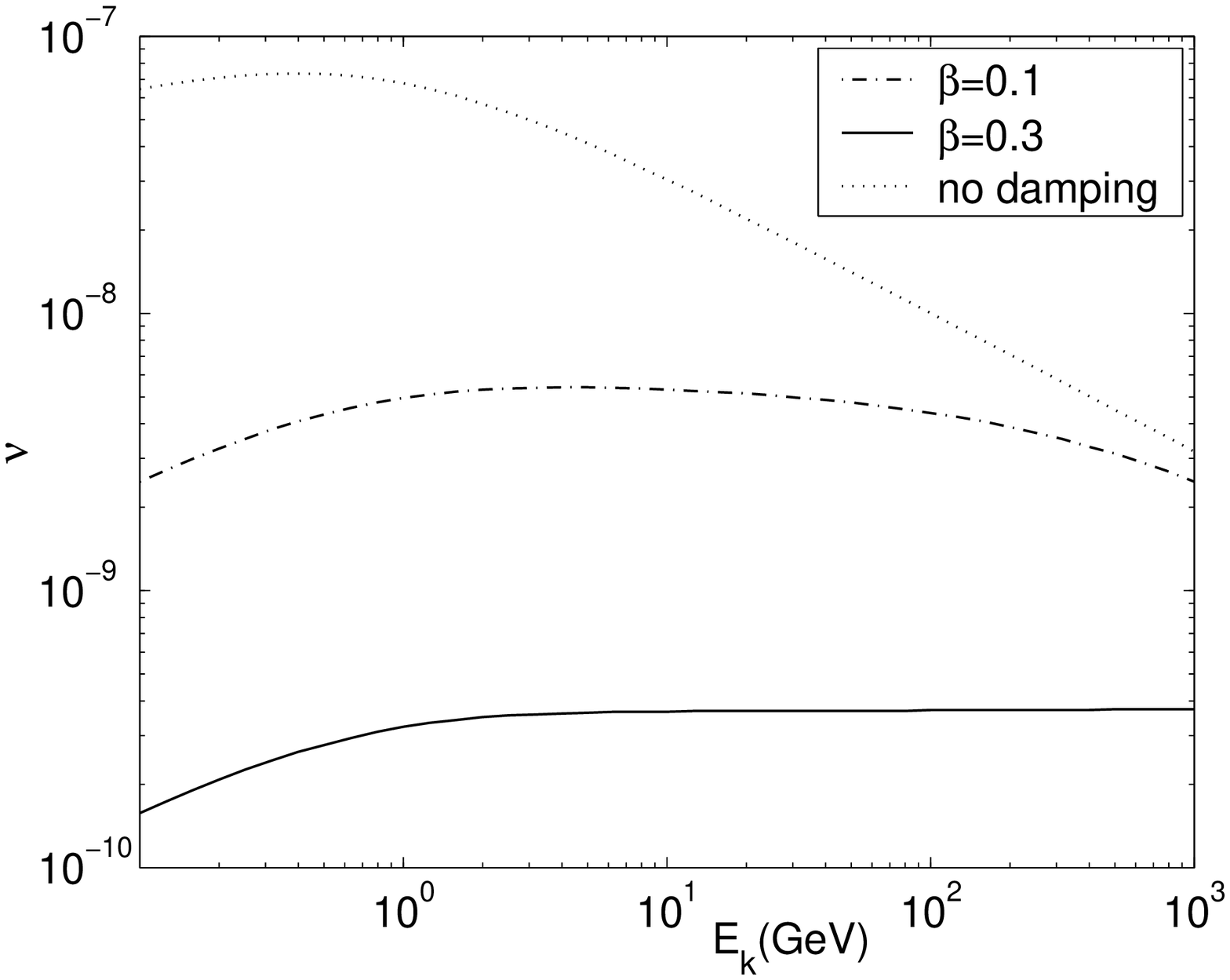}
} 
\caption{\small {\em Left:} rate of CR scattering by
Alfv\'en waves versus CR energy.  The lines at the top of the figure are
the accepted estimates obtained for Kolmogorov turbulence. The dotted
curve is from \cite{Chandran00}. The analytical calculations are given
by the solid line with our numerical calculations given by
crosses;  {\em Right:} the scattering by fast modes,  dashed line represents the case without damping for fast modes included, the solid and dash-dot line are the results taking into account collisionless damping.}
\label{impl}
\end{figure*}

Different from Alfv\'en and slow modes, fast modes are isotropic \citep{CL02_PRL}. Indeed they are subject to both collisional and collisionless damping. The studies in \cite{YL02, YL04} demonstrated, nevertheless, that the scattering by fast modes dominates in most cases in spite of the damping\footnote{On the basis of weak turbulence theory, \cite{Chandran2005} has argued that high-frequency 
fast waves, which move mostly parallel to magnetic field, generate Alfv\'en waves also moving mostly parallel to magnetic field. We expect
that the scattering by thus generated Alfv\'en modes to be similar to the scattering by the fast modes created by them. Therefore
we expect that the simplified approach adopted in \cite{YL04} and the papers that followed to hold.} (see Fig.\ref{impl} {\em right}).
More recent studies of cosmic ray propagation and acceleration that explicitly appeal to the effect of
the fast modes include \citet{Cassano_Brunetti, Brunetti_Laz, YL08, YLP08}.
Incidentally, fast modes have been also identified as primary agents for the acceleration of charged dust particles \cite{YL03,YLD04}.

\subsection{Nonlinear theory of diffusion}
\label{NLT_sec}

While QLT allows easily to treat the CR dynamics in a local magnetic
field system of reference, a key assumption in QLT, that the particle's orbit is unperturbed, makes one wonder about the limitations of the approximation. Indeed, while QLT provides simple physical insights into scattering, it is known to have problems. For instance, it fails in treating $90^\circ$ scattering  \citep[see][]{Volk:1973, Volk:1975, Jones:1973, Jones:1978, Owens:1974, Goldstein:1976, Felice90degree} and perpendicular transport \citep[see][]{Kota_Jok2000, Matthaeus:2003}. 

Indeed, many attempts have been made to improve the QLT and various non-linear
 theories have been attempted (see \citealt{Dupree:1966}, V\"olk 1973, 1975, 
Jones, Kaiser \& Birmingham 1973, Goldstein 1976). Currently we observe a surge
of interest in finding way to go beyond QLT. Examples include the nonlinear guiding center theory \citep[see][]{Matthaeus:2003}, second-order 
quasilinear theory \citep{Shalchi_SQT, Qin_NLT, LeRoux:2007}, etc. Most of the analysis were limited to traditional 2D+slab models of MHD turbulence. An important step was taken in Yan \& Lazarian (2008), where non-linear effect was accounted for in treating CR scattering in the type of MHD turbulence that are supported by numerical simulations. The results have been applied to both solar flares (Yan, Lazarian \& Petrosian 2008) and grain acceleration \citep{HLS12}. Below, we introduce the nonlinear theory and their applications to both particle transport and acceleration in incompressible and compressible turbulence based on the results from Yan \& Lazarian (2008).

The basic assumption of the quasi-linear theory is that particles follow unperturbed orbits. In reality, particle's pitch angle varies gradually with the variation of the magnetic field due to conservation of adiabatic invariant $v_\bot^2/B$, where $B$ is the total strength of the magnetic field \citep[see][]{Landau:1975}. Since B is varying in turbulent field, so are the projections of the particle speed $v_\bot$ and $v_\|$.
 This results in broadening of the resonance. The variation of the velocity is mainly caused by the magnetic perturbation $\delta B_\|$ in the parallel direction. This is true even for the incompressible turbulence we discussion in this section. For the incompressible turbulence, the parallel perturbation arises from the pseudo-Alfv\'en modes. The perpendicular perturbation $\delta B_\bot$ is higher order effect, which we shall neglect here.

The propagation of a CR can be described as a combination of a motion of its guiding center and CR's motion about its guiding center. 
Because of the dispersion of the pitch angle $\Delta\mu$ and therefore of the parallel speed $\Delta v_\|$, the guiding center is perturbed about the mean position $<z>=v\mu t$ as they move along the field lines. As a result, the perturbation $\delta B({\bf x},t)$  that the CRs view when moving along the field gets a different time dependence. The characteristic phase function $e^{ik_\|z(t)}$ of the perturbation $\delta B({\bf x},t)$ deviates from that for plane waves. Assuming the guiding center has a Gaussian distribution along the field line, \be
f(z)=\frac{1}{\sqrt{2\pi}\sigma_z}e^{-\frac{(z-<z>)^2}{2\sigma_z^2}},
\label{gauss}
\ee
one gets by integrating over z, \be
\int_{-\infty}^{\infty} dze^{ik_\| z}f(z)= e^{ik_\|<z>}e^{-k_\|^2\sigma_z^2/2}. 
\label{phase}
\ee
The first adiabatic invariant gives us 
\be
\sigma_z^2=<\Delta v_\|^2>t^2=\frac{v^4}{v_\|^2}\left(\frac{<\delta B_\parallel^2>}{B_0^2}\right)t^2.
\ee

Insert the Eq.(\ref{phase}) into the expression of $D_{\mu\mu}$ (see V\"olk 1975, \citealt{YL04}), we obtain

\bea
D_{\mu\mu}&=&\frac{\Omega^2(1-\mu^2)}{B_0^2}\int d^3k\sum_{n=0}^{\infty}R_n(k_{\parallel}v_{\parallel}-\omega\pm n\Omega)\nonumber\\
&&\left[I^A({\bf k})\frac{n^2J_n^2(w)}{w^2}+\frac{k_\|^2}{k^2}J^{'2}_n(w)I^M({\bf k})\right],
\label{general}
\eea 
Following are the definitions of the parameters in the above equation. $\Omega, \mu$ are the Larmor frequency and pitch angle cosine of the CRs. $J_n$ represents Bessel function, and $w=k_\bot v_\bot/\Omega=k_\bot LR\sqrt{1-\mu^2}$, where $R=v/(\Omega l)$ is the dimensionless rigidity of the CRs, $L$ is the injection scale of the turbulence. $k_\bot, k_\|$ are the components of the wave vector ${\bf k}$ perpendicular and parallel to the mean magnetic field, $\omega$ is the wave frequency. $I^A({\bf k})$ is the energy spectrum of the Alfv\'en modes and $I^M({\bf k})$ represents the energy spectrum of magnetosonic modes. In QLT, the resonance function $R_n=\pi\delta(k_{\parallel}v_{\parallel}-\omega\pm n\Omega)$. Now due to the perturbation of the orbit, it should be   
\bea
&&R_n(k_{\parallel}v_{\parallel}-\omega\pm n\Omega)\nonumber\\
&=&\Re\int_0^\infty dt e^{i(k_\|v_\|+n\Omega-\omega) t-\frac{1}{2}k_\|^2<\Delta v_\|^2>t^2}\nonumber\\
&=&\frac{\sqrt{\pi}}{|k_\|\Delta v_\||}\exp\left[-\frac{(k_\|v \mu-\omega+n\Omega)^2}{k_\|^2\Delta v_\|^2}\right]\nonumber\\
&\simeq&\frac{\sqrt{\pi}}{|k_\||v_\bot \sqrt{M_A}}\exp\left[-\frac{(k_\|v \mu-\omega+n\Omega)^2}{k_\|^2v_\bot^2M_A}\right]
\label{resfunc}
\eea
where $M_A\equiv \delta V/v_A=\delta B/B_0$ is the Alfv\'enic Mach number and $v_A$ is the Alfv\'en speed. We stress that Eqs.~(\ref{general},\ref{resfunc}) are generic, and applicable to both incompressible and compressible medium. 

For gyroresonance ($n=\pm 1,2,...$), the result is similar to that from QLT for $\mu\gg \Delta \mu=\Delta v_\|/v$. In this limit, Eq.(\ref{general}) represents a sharp resonance and becomes equivalent to a $\delta$-function when put into Eq.(\ref{general}).  
In general, the result is different from that of QLT, especially at $\alpha\rightarrow 90^\circ$, the resonance peak happens at $k_{\|,res}\sim \Omega/\Delta v$ in contrast to the QLT result 
$k_{\|,res}\sim\Omega/v_\|\rightarrow \infty$. We shall
show below, that due to the anisotropy, the scattering coefficient $D_{\mu\mu}$ is still very small if the Alfv\'en and the pseudo-Alfv\'en modes are concerned. 

On the other hand, the dispersion of the $v_\parallel$ means that CRs with a much wider range of pitch angle can be scattered by the compressible modes through TTD 
($n=0$), which is marginally affected by the anisotropy and much more efficient than the gyroresonance. In QLT, the projected particle speed should be comparable to phase speed of the magnetic field compression according to the $\delta$ function for the TTD resonance.. This means that only particles with a specific pitch angle 
can be scattered. For the rest of the pitch angles, the interaction is still dominated by gyroresonance, which efficiency is negligibly small for the Alfv\'enic anisotropic turbulence (see \S\ref{Alf_scatter}). With the resonance broadening, however, wider range of pitch angle can be scattered through TTD, including $90^\circ$. 


\subsection{Results from test particle simulations}

We live in an era when we can test various processes in astrophysics and numerical studies have become an important part of theoretical efforts. Test particle simulation has been used to study CR scattering and
transport \cite{Giacalone_Jok1999, Mace2000}. The aforementioned studies, however, used synthetic
data for turbulent fields, which have several disadvantages.
Creating synthetic turbulence data which has scale-dependent
anisotropy with respect to the local magnetic field (as observed
in \citealt{CV00} and \citealt{MG01}) is difficult
and has not been realized yet.  Also,
synthetic data normally uses Gaussian statistics and
delta-correlated fields, which is hardly appropriate for
description of strong turbulence. 

\begin{figure}
\includegraphics[width=0.45\textwidth]{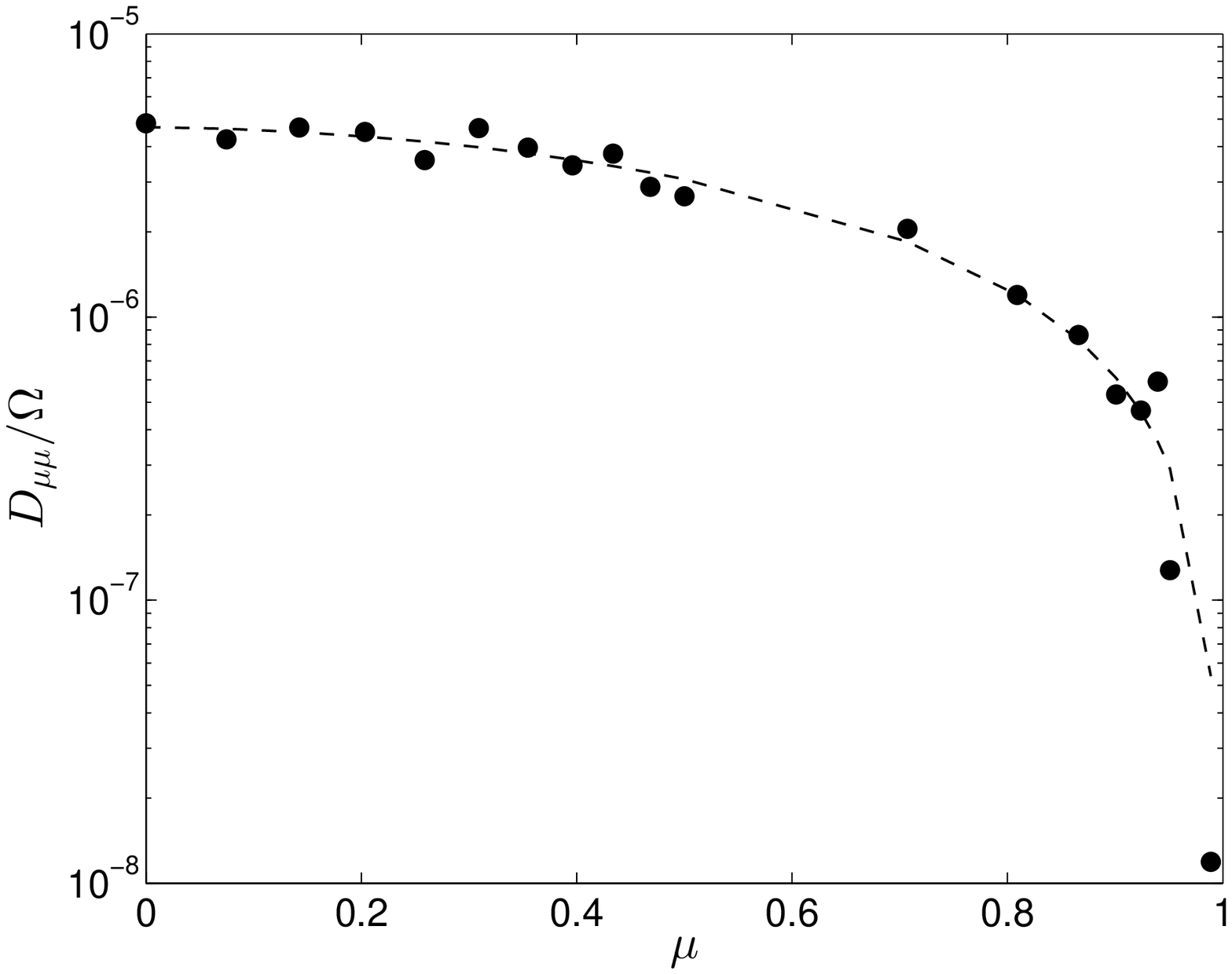}
\includegraphics[width=0.45\textwidth]{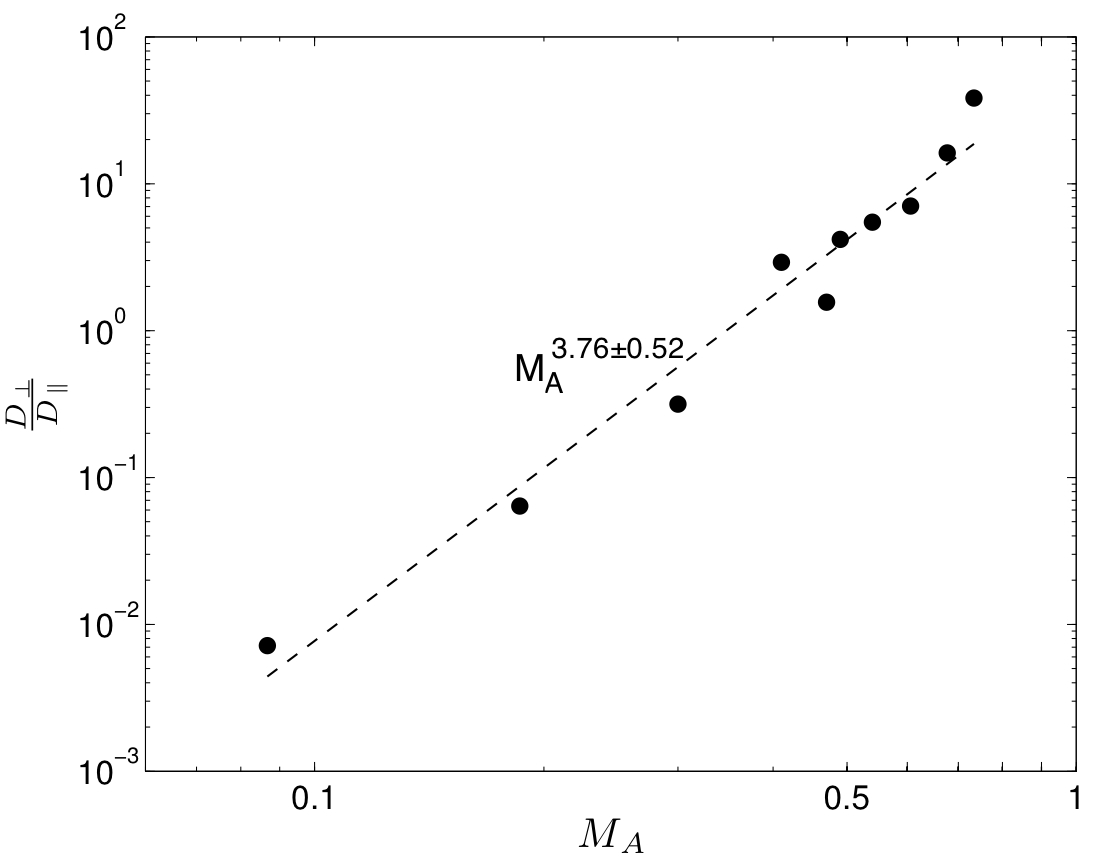}
\caption{{\em Left}: dimensionless CR scattering coefficient $D_{\mu
    \mu}/\Omega$ vs the pitch angle $\mu$. It is dominated by TTD resonant mirror interaction with compressible modes; {\em right:} Diffusive behavior of the particles displayed in the tracing
  simulations.  Both the parallel and perpendicular transport are normal diffusion, and the ratio of their diffusion coefficients is $\sim M_A^4$, consistent with the analytical prediction in \cite{YL08} \citep*[from][]{Xu_Yan}.}
\label{xx_yy}
\end{figure}

Using the results of direct numerical MHD simulations as the input data, \cite{BYL2011} and \cite{Xu_Yan} performed test particle simulations. Their results show good correspondence with the analytical predictions. We briefly summarize the results here. As shown in Fig.\ref{xx_yy}, particles' motion is diffusive both along the magnetic field (x direction) and across the field (y direction). Moreover, the scattering coefficient shows the same pitch angle dependence as that predicted in \cite{YL08}, namely the scattering is most efficient for large pitch angles due to the TTD mirror interaction (see Fig. \ref{xx_yy} {\em left}).  

\section{Cosmic ray propagation in Galaxy}
\label{results}
The scattering by fast modes is influenced by the medium properties as the fast modes are subject to linear damping, e.g., Landau damping.
 Using the approach above we revisit the problem of the CR propagation in the selected phases of the ISM (see Table~\ref{ch1t1} for a list of fiducial parameters appropriate for the idealized phases\footnote{The parameters of idealized interstellar phases are a subject of debate. Recently, even the entire concept of the phase being stable
 entities has been challenged \citep[see][and ref. therein]{Gazol:2007}. Indeed different parts
 of interstellar medium can exhibit variations of these parameters \citep[see][and ref. therein]{Wolfire:2003}}) assuming that turbulence is injected on large scales.
\begin{table*}
{\footnotesize \begin{tabular}{ccccccc}
\hline
\hline 
 ISM&
halo&
 HIM&
 WIM&
 WNM&
 CNM&
 DC\tabularnewline
\hline
T(K)&
 $2\times 10^6$&
 $1\times10^{6}$&
 8000&
 6000&
 100&
 15\tabularnewline
$c_S$(km/s)&
130&
91&
8.1&
7&
0.91&
0.35\tabularnewline
n(cm$^{-3}$)&
 $10^{-3}$&
 $4\times10^{-3}$&
 0.1&
 0.4&
 30&
 200\tabularnewline
$l_{mfp}$(cm)&
$4\times 10^{19}$&
$2\times10^{18}$&
$6\times10^{12}$&
$8\times10^{11}$&
$3\times10^{6}$&
$10^{4}$\tabularnewline
L(pc)&
 100&
 100&
 50&
 50&
 50&
 50\tabularnewline
B($\mu$G)&
5&
2&
5&
5&
5&
15\tabularnewline
$\beta$&
0.28&
3.5&
0.11&
0.33&
0.42&
0.046\tabularnewline
damping&
 collisionless&
 collisional&
 collisional&
 neutral-ion&
 neutral-ion&
 neutral-ion\tabularnewline
\hline
\hline
\end{tabular}
\caption{The parameters of idealized ISM phases and relevant damping. The
dominant damping mechanism for turbulence is given in the last line. HIM=hot ionized medium, CNM=cold neutral medium, WNM=warm neutral
medium, WIM=warm ionized medium, DC=dark cloud.}}
\label{ch1t1}
\end{table*}

\subsection{Halo}

\begin{figure}
\includegraphics[width=0.45\textwidth]{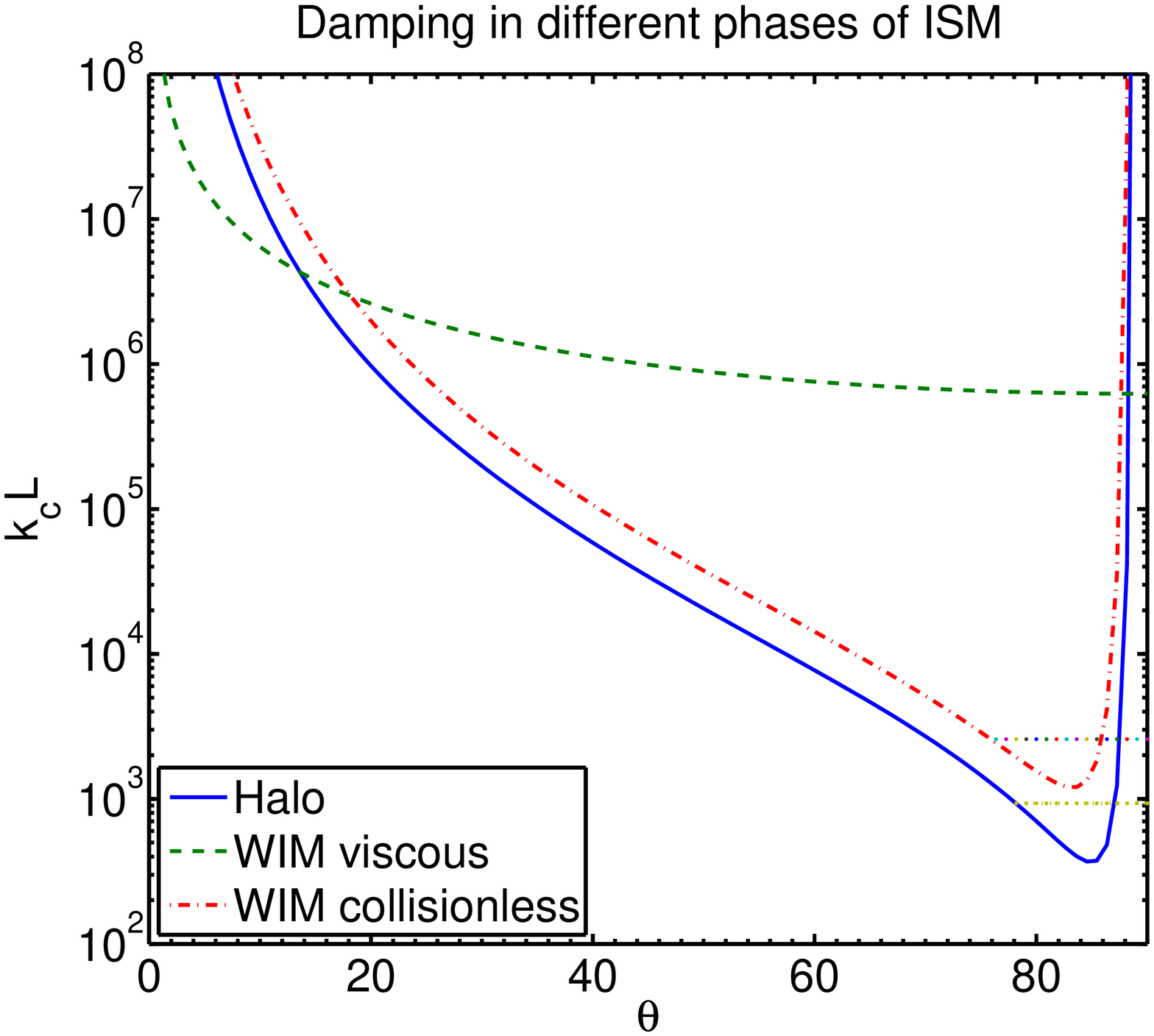}
\includegraphics[width=0.45\textwidth]{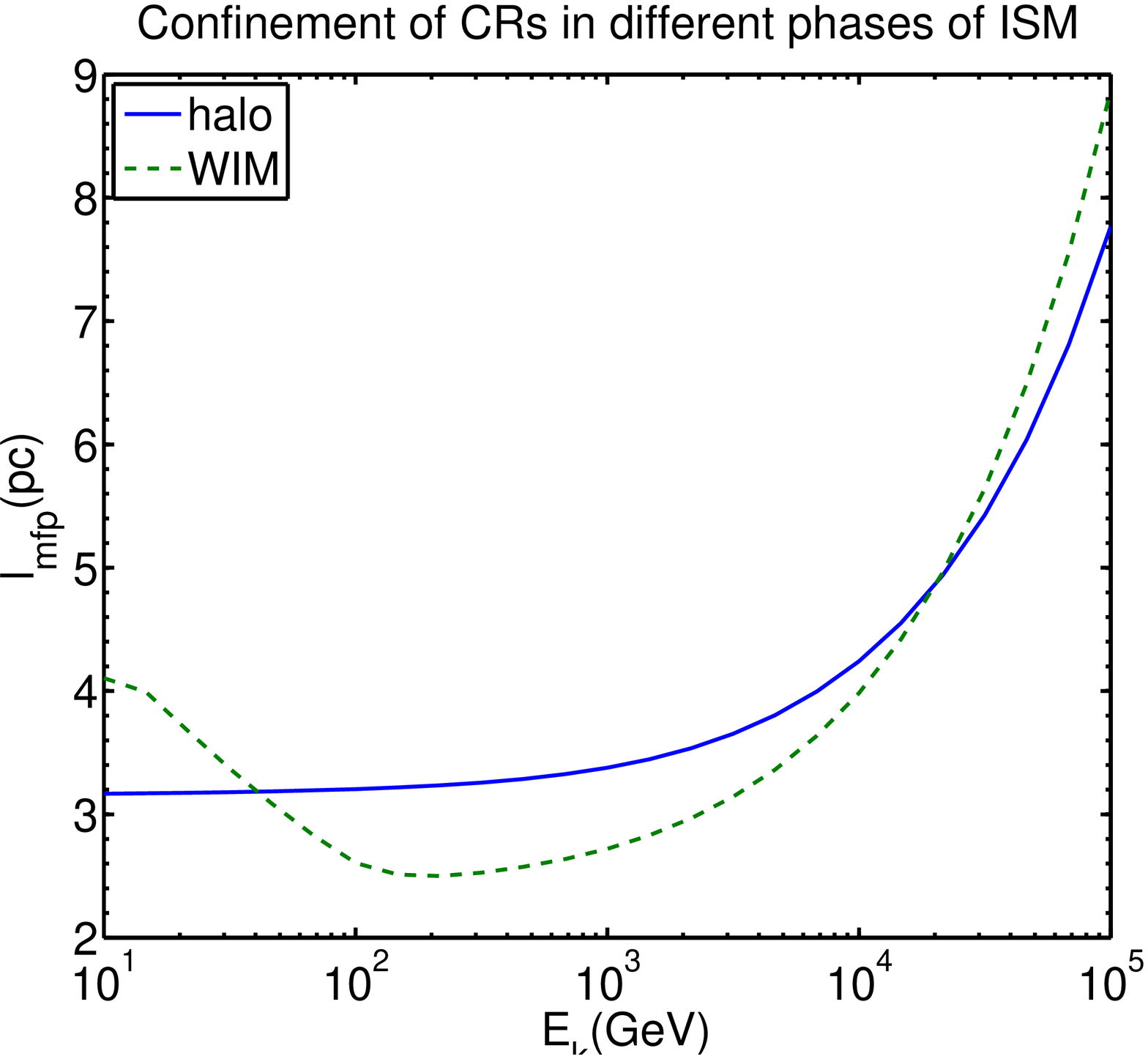}
\caption{\small {\em Left}: The turbulence truncation scales in Galactic halo and warm ionized medium (WIM). The damping curves flattens around $90^\circ$ due to field line wandering (dotted lines, see \citealt*{YL04, LVC04}); For WIM, both viscous and collisionless damping are applicable; {\em right}: The mean free paths in two different phases of ISM: halo (solid line) and WIM (dashed line). At lower energies ($\sim<100$GeV), the different dependence in WIM is owing to the viscous damping \citep[from][]{YL08}.}
\label{mfp}
\end{figure}

In Galactic halo (see Table~\ref{ch1t1}), the Coulomb collisional mean free path is $\sim 10$ pc, the plasma is thus in a collisionless regime. The cascading rate 
of the fast modes is \citep{CL02_PRL}
\be
\tau_k^{-1}=(k/L)^{1/2}\delta V^2/V_{ph}.
\label{tcasfast}
\ee

By equating it with the collisionless damping rate 
\be
\Gamma_{c} = \frac{\sqrt{\pi\beta}\sin^{2}\theta}{2\cos\theta}kv_A\times \left[\sqrt{\frac{m_e}{m_i}}\exp\left(-\frac{m_e}{\beta m_i\cos^2\theta}\right)+5\exp\left(-\frac{1}{\beta\cos^{2}\theta}\right)\right],
\label{Ginz}
\ee
we obtain the turbulence truncation scale $k_c$:
\be
k_c L\simeq \frac{4M_A^4m_i\cos^2\theta}{\pi m_e\beta\sin^4\theta}\exp\left(\frac{2m_e}{\beta m_i\cos^2\theta}\right).
\label{landauk}
\ee
where $\beta=P_{gas}/P_{mag}$.

The scale $k_c$ depends on the {\it wave pitch angle} $\theta$, which makes
the damping anisotropic. As the turbulence undergoes cascade and the waves propagate in a turbulent medium, the angle $\theta$ is changing.
As discussed in YL04 the field wandering defines the spread of angles. During one cascading time, the fast modes propagate a distance 
$v\tau_{cas} $ and see an angular deviation $\tan \delta \theta \simeq \sqrt{\tan^2\delta \theta_\parallel+\tan^2 \delta\theta_\perp}$, which is
\be
\tan \delta\theta \simeq  \sqrt{\frac{M_A^2\cos\theta}{27(kL)^{1/2}}+\left(\frac{M_A^2\sin^2\theta}{kL}\right)^{1/3}}
\label{dthetaB}
\ee
As evident, the damping scale given by Eq.(\ref{landauk}) varies considerably especially when $\theta\rightarrow 0$ and $\theta\rightarrow 90^\circ$. For the quasi-parallel modes, the randomization ($\propto (kL)^{-1/4}$) is negligible since the turbulence cascade continues to very small scales. On small scales, most energy of the fast modes is contained in these quasi-parallel modes \citep*{YL04, Petrosian:2006}.

For the quasi-perpendicular modes, the damping rate (Eq.\ref{Ginz}) should be averaged over the range $90^\circ-\delta\theta$ to $90^\circ$. Equating Eq.(\ref{tcasfast}) and Eq.(\ref{Ginz}) averaged over $\delta\theta$, we get the averaged damping wave number (see Fig.\ref{mfp} {\em left}). The field line wandering has a marginal effect on the gyroresonance, whose interaction with the quasi-perpendicular modes is negligible (YL04). However, TTD scattering rates of moderate energy CRs ($<10$TeV) will be decreased owing to the increase of the damping around the $90^\circ$ (see Fig.\ref{mfp} {\em left}). For higher energy CRs, the influence of damping is marginal and so is that of field line wandering. 

\begin{figure}
\includegraphics[width=0.45\textwidth]{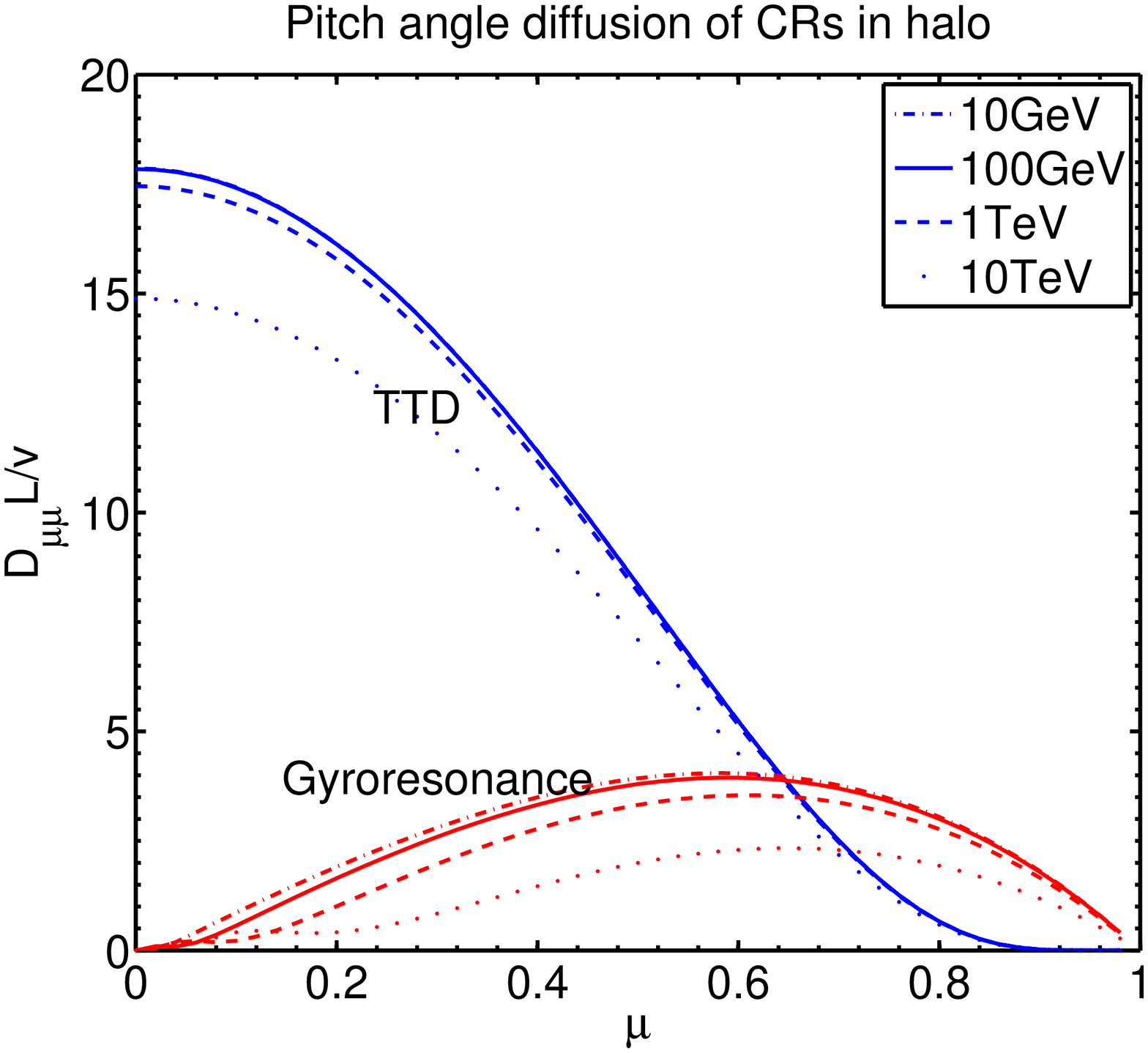}
\includegraphics[width=0.45\textwidth]{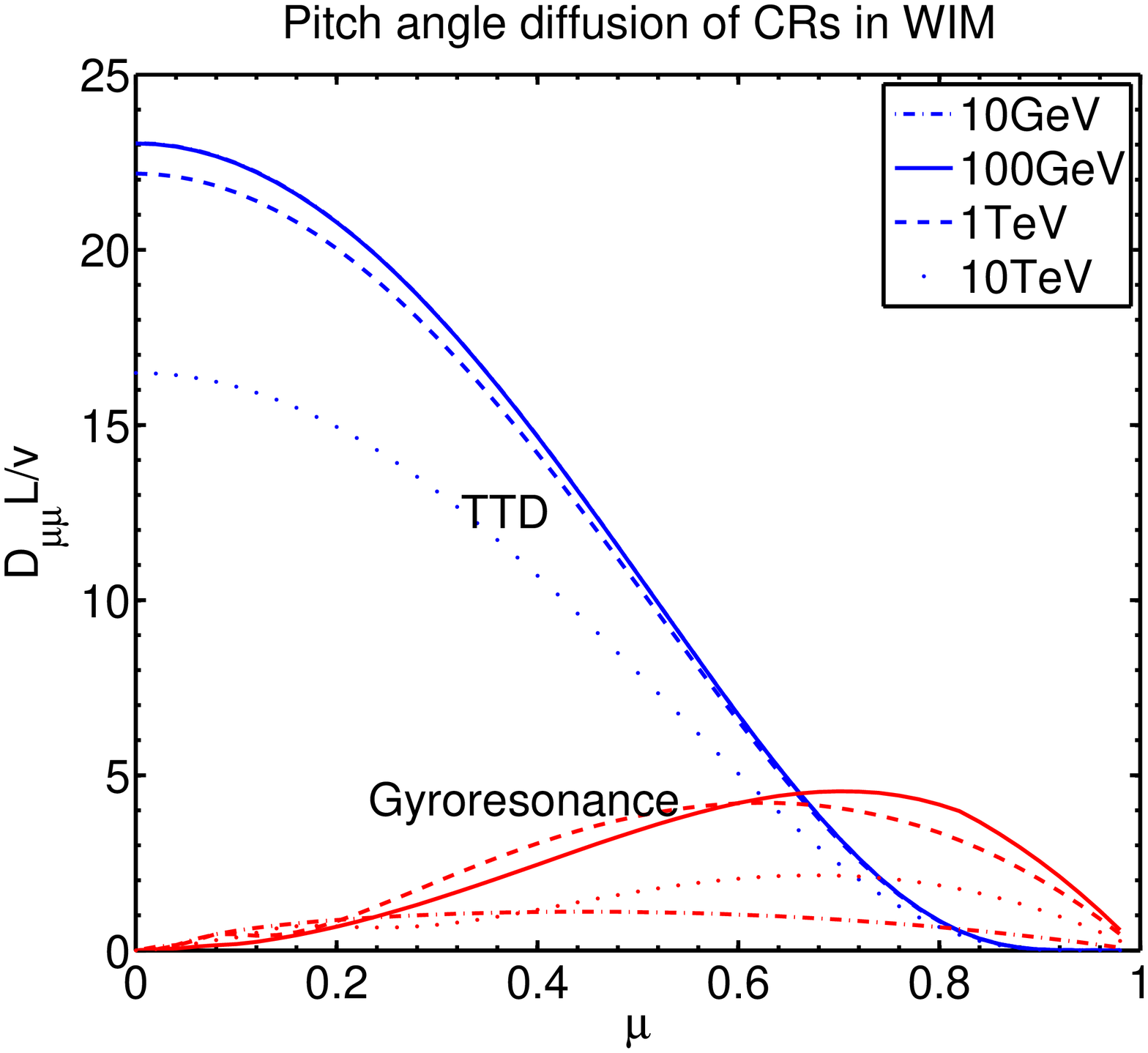}
\caption{Pitch angle diffusion coefficients in halo and WIM. Upper lines in the plots represent the contribution from TTD and lower lines are for gyroresonance \citep[from][]{YL08}.}
\label{fastcompr}
\end{figure}

The QLT result on gyroresonance in the range $\mu>\Delta \mu$ provides a good approximation to the non-linear results \citep{YL08}. For CRs with sufficiently small rigidities, the resonant fast modes ($k_{res}\approx 1/(R\mu)$) are on small scales with a quasi-slab structure (see Fig.\ref{mfp} {\em left}). For the scattering by these quasi-parallel modes, the analytical result that follows from QLT approximation \citep[see][]{YL04} for the gyroresonance is\footnote{It can be shown that the QLT result follows from our more general results (see Eqs.\ref{general}, \ref{resfunc}) if we put $\Delta \mu \rightarrow 0$. This justifies our use of the analytical approximation.} 

\be
\left[\begin{array}{c}
D^{G}_{\mu\mu}\\
D^{G}_{pp}\end{array}\right]=\frac{\pi v \mu^{0.5}(1-\mu^{2})}{4LR^{0.5}}\left[\begin{array}{c}
\frac{1}{7}[1+(R\mu)^2]^{-\frac{7}{4}}-(\tan^{2}\theta_c+1)^{-\frac{7}{4}}\\
\frac{m^{2}V_{A}^{2}}{3}\left\{[1+(R\mu)^2]^{-\frac{3}{4}}-(\tan^{2}\theta_c+1)^{-\frac{3}{4}}\right\}\end{array}\right]
\label{lbgyro}\ee
where $\tan\theta_c={k_{\perp,c}}/{k_{\parallel,res}}$.

Once we know the functional form of the $D_{\mu\mu}$, we can obtain the corresponding mean free path \citep{Earl:1974}:
\be
\lambda_\|/L=\frac{3}{4}\int^1_0 d\mu \frac{v(1-\mu^2)^2}{(D^T_{\mu\mu}+D^G_{\mu\mu})L},
\ee
where $D^T_{\mu\mu}$ is the contribution from TTD interaction and can be obtained using the nonlinear theory (see \citealt{YL08}, and also \S\ref{NLT_sec}) with the inertial range of fast modes determined for the local medium (see, e.g. \ref{landauk} in the case of collisionless damping). 

The mean free path is sensitive to the scattering by gyroresonance at small pitch angles, due to the influence of damping on the fast modes on small scales. Fig.\ref{fastcompr} shows the pitch angle diffusion of CRs with different energies due to the TTD and gyroresonance.
 
The weak dependence of the mean free path (see Fig.\ref{mfp} {\em right}) of the moderate energy (e.g$<1$TeV) CRs in halo results from the fact that gyroresonance changes marginally with the CR energy (see Fig.\ref{fastcompr}). This is associated with the damping in collisionless medium. We expect that similar flat dependence can happen in any collisionless medium. This can be a natural explanation of the puzzling ``Palmer Concensus" \citep{Palmer:1982}, the same trend observed in solar wind.

\subsection{Warm Ionized Medium}

In warm ionized medium, the Coulomb collisional mean free path is $l_{mfp}=6\times 10^{12}$ cm and the plasma $\beta\simeq0.11$. Suppose that the turbulence energy is injected from large scale, then the compressible turbulence is subjected to the viscous damping besides the collisionless damping. 
By equating the viscous damping rate with the cascading rate (Eq.\ref{tcasfast}), we obtain the following truncation scale, 
\bea
k_{c}L=x_c\left\{\begin{array}{rl}(1-\xi^2)^{-\frac{2}{3}} & \beta\ll 1\\
(1-3\xi^2)^{-\frac{4}{3}} & \beta\gg 1\end{array}\right.
\eea
where $x_c=\left[\frac{6\rho\delta V^2L}{\eta_0V_A}\right]^{\frac{2}{3}}$,
 $\eta_0$ is the longitudinal viscosity. In the low $\beta$ regime, the motions are primarily perpendicular to the magnetic field so that $\partial v_{x}/\partial x=\dot{n}/n\sim\dot{B}/B$. The longitudinal viscosity enters here as the result of distortion of the Maxiwellian distribution \citep[see][]{Braginskii:1965}. The transverse energy of the ions increases during compression because of the conservation of adiabatic invariant $v_{\perp}^{2}/B$. If the rate of compression is faster than that of collisions, the ion distribution in the momentum space is bound to be distorted from the Maxiwellian isotropic sphere to an oblate spheroid with the long axis perpendicular to the magnetic field. As a result, the transverse pressure gets greater than the longitudinal pressure, resulting in a stress $\sim\eta_{0}\partial v_{x}/\partial x$.
The restoration of the equilibrium increases the entropy and causes the dissipation of energy.

The viscous damping scale is compared to collisionless cutoff scale (Eq.\ref{landauk}) in Fig.\ref{mfp} {\it left}. 
As shown there, both viscous damping and collisionless damping are important in WIM. Viscous damping is dominant for small $\theta$ and
 collisionless damping takes over for large $\theta$ except for $\theta=90^\circ$.
This is because collisionless damping increases with $\theta$ much faster than the viscous damping. For sufficiently small wave pitch angles, the viscous damping is too small to prevent the fast modes to cascade down to scales smaller than the mean free path $l_{mfp}$. Because of the similar quasi-slab structure on small scales, 
Eq.(\ref{lbgyro}) can be also applied in WIM. The results are illustrated in Fig.\ref{fastcompr}. Compared to the case in halo, we see that the qualitative difference stands in the gyroresonance. This is because gyroresonance is sensitive to the quasi-slab modes whose damping differs in halo and WIM.  

\subsection{Other phases}

In hot ionized medium (HIM), the plasma is also in collisionless regime, but the density is higher and the plasma beta is larger than 1. The damping by protons thus becomes substantial especially at small pitch angles. The damping truncates the turbulence at much larger scales than the gyroscales of the CRs of the energy range we consider. No gyroresonance can happen and some other mechanisms are necessary to prevent CRs streaming freely along the field. The turbulence injected from small scales might play an important role (see \S6).

In partially ionized gas one should take into account an additional damping that arises from ion-neutral collisions \citep[see][]{Kulsrud_Pearce, LG01, LVC04}. In the latter work a viscosity-damped regime of turbulence was predicted at scales less the scale $k_{c, amb}^{-1}$ at which the ordinary magnetic turbulence is damped by ionic viscosity. The corresponding numerical work, e.g., \cite{CLV_newregime} testifies that for the viscosity-damped regime the parallel scale stays equal to the scale of the ambipolar damping, 
i.e., $k_{\|}=k_{c, amb}$, while $k_{\bot}$ increases. In that respect, the scattering by such magnetic fluctuations is analogous to the scattering induced by the weak turbulence (see \S 2.3, \citealt{YL08}). The difference stems from the spectrum
of $k_{\bot}$ is shallower than the spectrum of the weak turbulence. The predicted values of the spectrum for the viscosity-damped turbulence $E(k_\bot)\sim
k_\bot^{-1}$ \citep{LVC04} are in rough agreement with simulations. More detailed studies of scattering in partially ionized gas will be necessary.    

\section{Perpendicular transport}

In this section we deal with the diffusion perpendicular
to {\it mean} magnetic field. 

Propagation of CRs perpendicular to the mean magnetic field is another important problem in which QLT encounters serious difficulties.
Compound diffusion, resulting from the convolution of diffusion along the magnetic field line and diffusion of field line perpendicular to mean field direction, has been invoked to discuss transport of cosmic rays in the Milky Way \citep*{Getmantsev, Lingenfelter:1971,Allan:1972}. The role of compound diffusion in the acceleration
 of CRs at quasi-perpendicular shocks were investigated by \cite{Duffy:1995} and \cite{Kirk:1996}.  

Indeed, the idea of CR transport in the direction perpendicular to the mean magnetic field being dominated by the field line random walk 
(FLRW, \citealt{Jokipii1966, Jokipii_Parker1969, Forman1974}) can be easily justified
only in a restricted situation where the turbulence perturbations are small and CRs do not scatter backwards to retrace their trajectories. If the latter is not true, the particle motions are subdiffusive, 
i.e., the squared distance diffused growing as not as $t$ but as $t^{\alpha}$, $\alpha<1$, e.g., $\alpha=1/2$ \citep{Kota_Jok2000, Mace2000, Qin2002}.
If true, this could indicate a substantial shift in the paradigm of CR transport, a shift that surely dwarfs a modification of magnetic turbulence model from the 2D+slab to a more simulation-motivated model that we deal here.

It was also proposed that with substantial transverse structure, {\it i.e.}, transverse displacement of field lines, perpendicular diffusion is recovered \citep{Qin2002}. Is it the case of the MHD turbulence models we deal with?

How realistic is the subdiffusion in the presence of turbulence? The answer for this question apparently depends on the models of turbulence chosen. 

Compound diffusion happens when particles are restricted to the magnetic field lines and perpendicular transport is solely due to the random walk of field line wandering \citep[see][]{Kota_Jok2000}. 
In the three-dimensional turbulence, field lines are diverging away due to shearing by the Alfv\'en modes \citep[see][]{LV99, Narayan_Medv, Lazarian06}.
 Since the Larmor radii of CRs are much larger than the minimum scale of eddies $l_{\bot, min}$, field lines within the CR Larmor orbit are effectively diverging away owing to shear by the Alfv\'enic turbulence.
The cross-field transport thus results from the deviations of field lines at small scales, as well as field line random walk at large scale ($>{\rm min}[L/M^3_A,L]$).

Both observation of Galactic CRs and solar wind indicate that the diffusion of CRs perpendicular to magnetic field is normal diffusion \citep[]{Giacalone_Jok1999, Maclennan2001}. Why is that?

Most recently the diffusion in magnetic fields was considered for thermal particles in Lazarian (2006), for cosmic rays in \cite{YL08}. In what follows we present the results based on the studies in  \cite{YL08}.

\subsection{Perpendicular diffusion on large scale}

For perpendicular diffusion, the important issue is the reference frame. We emphasize that we consider the diffusion perpendicular to the {\emph mean} field direction in the global reference of frame.  

{\it High $M_A$ turbulence}: High $M_A$ turbulence corresponds to the field that is easily bended by
hydrodynamic motions at the injection scale as the hydro energy at the
injection scale is much larger than the magnetic energy, i.e.
$\rho V_L^2\gg B^2$. In this case
magnetic field becomes dynamically important on a much smaller scale, i.e. the 
scale $l_A=L/M_A^3$ \citep[see][]{Lazarian06}. If the parallel mean free path of CRs $\lambda_\|\ll l_A$, the stiffness of B field is negligible so that the perpendicular diffusion coefficient is the same as the parallel one, i.e., $D_\bot=D_\|\sim 1/3 \lambda_{\|} v$. If $\lambda_\|\gg l_A$, the
 diffusion is controlled by the straightness of the field lines, and $
D_\bot=D_{\|}\approx 1/3l_Av.
\label{dbb}
$ The diffusion is isotropic if scales larger than $l_A$ are
concerned. 

{\it Low $M_A$ turbulence}: In the magnetically dominated case, i.e. the field that cannot be easily bended at
the turbulence injection scale, individual magnetic field lines are aligned
with the mean magnetic field. The diffusion in this case is anisotropic.
If turbulence is injected at scale $L$ it stays 
weak for the scales larger than $LM_A^2$ and it is 
strong at smaller scales. Consider first the case of $\lambda_\|>L$.
The time of the individual step is $L/v_\|$, then $D_\perp\approx 1/3Lv M_A^4.$
This is similar to the case discussed in the FLRW model (Jokipii 1966). However, we obtain the dependence of $M_A^4$ instead of their $M_A^2$ scaling. In the opposite case of $\lambda_\|<L$, the perpendicular diffusion coefficient is $
D_{\bot}\approx D_{\|}M_A^4,
\label{diffx}$ which coincides with the result
obtained for the diffusion of thermal electrons in magnetized plasma \citep{Lazarian06}. This is due to the anisotropy of the Alfv\'enic turbulence.

\subsection{Superdiffusion on small scales}

The diffusion of CR on the scales $\ll L$ is different and it is determined by how fast field lines are diverging away from each other. The mean deviation of a field in a distance $\delta x$ is proportional to $[\delta z]^{3/2}$  \citep{LV99, Lazarian06}, same as Richardson diffusion in the case of hydrodynamic turbulence \cite[see][]{Eyink2011}. Following the argument, we showed in \cite{YL08} that the cosmic ray perpendicular transport is superdiffusive. The reason is that there is no random walk on small scales up to the injection scale of strong MHD turbulence ($LM_A^2$ for $M_A < 1$
 and $l_A$ for $M_A>1$). This can well explain the recently observed super-diffusion in solar wind \citep{Perri2009}. Superdiffusion can have important implications for shock acceleration as discussed in details in \cite{LY13}.

\subsection{Is there subdiffusion?}
The diffusion coefficient $D_{\|}M_A^4$ we obtained in the case of $M_A<1$, means that the transport
perpendicular to the dynamically strong magnetic field is a normal diffusion, rather
than the subdiffusion as discussed in a number of recent papers. This is also supported by test particle simulations (\citealt*{BYL2011, Xu_Yan}, see Fig.\ref{xx_yy} {\it right}). Let us
clarify this point by obtaining the necessary conditions for the subdiffusion
to take place.

The major implicit assumption in subdiffusion (or compound diffusion) is that the particles trace back 
their 
trajectories in x direction on the scale $\delta z$. When is it possible to talk about retracing of particles? In the case of random motions at a single scale {\it only}, the distance over 
which the particle
trajectories get uncorrelated is given by the \cite{RR1978}
model. Assuming that the damping scale of the turbulence  is larger
that the CR Larmor radius, the \cite{RR1978}
model, when generalized to anisotropic turbulence provides \citep{Narayan_Medv, Lazarian06} $L_{RR}=l_{\|, min}\ln(l_{\bot, min}/r_{Lar})$
where $l_{\|, min}$ is the parallel scale of the cut-off of turbulent motions, 
$l_{\bot, min}$ is the corresponding perpendicular scale, $r_{Lar}$ is the
CR Larmor radius. The assumption of $r_{Lar}<l_{\bot, min}$ can be
 valid, for instance, for Alfv\'enic motions in partially ionized gas.
However, it is easy to see that, even in this case,
 the corresponding scale is rather
small and therefore subdiffusion is not applicable for the transport
of particles in Alfv\'enic turbulence over scales $\gg l_{\|,min}$.

If  $r_{Lar}>l_{\bot, min}$, as it is a usual case for Alfv\'en motions in the
phase of ISM with the ionization larger than $\approx 93\%$, where the
Alfv\'enic motions go to the thermal particle gyroradius 
\citep[see estimates in][]{LG01, LVC04}, 
the subdiffusion of CR is not an applicable concept for Alfv\'enic turbulence.  
This does
not preclude subdiffusion from taking place
in particular models of magnetic perturbations,
e.g. in the slab model considered in \cite{Kota_Jok2000}, but we believe in the omnipresence of Alfv\'enic turbulence in interstellar gas \citep[see][]{Armstrong95}.

\section{Streaming Instability in the Presence of Turbulence}

\begin{table}
\caption{The notation we used in this section}
\label{notations}
\begin{tabular}{|c|r|}
\hline
A  & normalized wave amplitude $\delta B/B_0$\\
a& hardening of the CR spectrum at the shock front\\
$B_0$ & mean magnetic field at the shock in the later Sedov phase\\
$B_{cav}$ &inercloud magnetic field strength\\
$\delta B$& wave amplitude\\
c & light speed\\
d& distance of the molecular cloud from observer\\
D& diffusion coefficient of CRs\\
E& CR energy\\
$E_{SN}$& supernova explosion\\
f& distribution function of CRs\\
$f_\pi$& fraction of energy transferred from parent protons to pions\\
k& wave number\\
K(t) & Normalization factor of CR distribution function\\
L& the injection scale of background turbulence\\
m& proton rest mass\\
$M_c$ & cloud mass\\ 
n& intercloud number density\\ 
$N_\gamma$ & $\gamma$ ray flux\\
p& CR's momentum\\
$p_{max}$& the maximum momentum accelerated at the shock front\\
$P_{CR}$ & CR pressure\\
q & charge of the particle\\
r& distance from SNR centre\\
$R_c$& the distance of the molecular cloud from the SNR centre\\
$r_g$ & Larmor radius of CRs\\
$R_d$ & diffusion distance of CRs\\
$R_{sh}$ & shock radius\\
$R_{esp},\,t_{esp}$& the escaping distance/time of CRs\\ 
s& 1D spectrum index of CR distribution\\
t& time since supernova explosion\\
$t_{age}$ & the age of SNR\\
$t_{sed}$ & the time at which SNR enters the Sedov phase\\
U& shock  speed\\
$U_i$ &initial shock velocity\\
v& particle speed\\
$v_s$ & streaming speed of CRs\\
W& wave energy\\
$\alpha$& power index of D with respect to particle momentum p\\
$\chi$&reduction factor of D with respect to $D_{ISM}$\\
$\delta$& power index of $p_{max}$ with respect to t\\
 $\eta$&fraction of SN energy converted into CRs\\
 $\eta_A$& a numerical factor in Eq.\ref{pp}\\
$\Gamma_{cr}$& the growth rate of streaming instability\\
$\Gamma_d$& wave damping rate\\
 $ \kappa$&ratio of diffusion length to shock radius\\
 $\Omega_0$ & the Larmor frequency of non-relativistic protons\\
 $\sigma_{pp}$ &cross section for pp collision\\
 $\xi$& the ratio of CR pressure to fluid ram pressure\\
\hline
\end{tabular}
\end{table}
                
When cosmic rays stream at a velocity much larger than Alfv\'{e}n
velocity, they can excite by gyroresonance MHD modes which in turn scatter
cosmic rays back, thus increasing the amplitude of the resonant mode. This 
runaway process is known as streaming instability. It was claimed
that the instability could provide confinement for cosmic rays with
energy less than $\sim 10^2$GeV \citep{Cesarsky80}. However, this was calculated
in an ideal regime, namely, there was no background MHD turbulence.
In other words, it was thought that the
self-excited modes would not be appreciably damped in fully ionized gas\footnote{We neglect the nonlinear Landau damping, which is suppressed in turbulence due to decrease of mean free path.}. 

This is not true for turbulent medium, however. \citet{YL02}
pointed out that the streaming instability is partially suppressed
in the presence of background turbulence \citep[see more in][]{LCY02}. More recently, detailed calculations of the 
streaming instability in the presence of background Alfv\'enic turbulence 
were presented in \cite{FG04}. The growth rate of the modes of wave number $k$ is \citep{Longairbook}.
\begin{equation}
\Gamma_{cr}(k)=\Omega_0\frac{N(\geq E)}{n_{p}}(-1+\frac{v_{stream}}{V_{A}}),
\label{instability}
\end{equation}
 where $N(\geq E)$ is the number density of cosmic rays with energy
$\geq E$ which resonate with the wave,
$n_{p}$ is the number density of charged particles in the medium.
The number density of cosmic rays near the sun is $N(\geq E)\simeq 2\times10^{-10}(E/$GeV$)^{-1.6}$
cm$^{-3}$sr$^{-1}$ \citep{Wentzel74}.

Interaction with fast modes was considered by \citet{YL04}. 
Such an interaction happens at the rate 
$\tau_k\sim (k/L)^{-1/2}V_{ph}/V^2$. 
By equating it with the growth rate Eq.(\ref{instability}), we can find that the streaming instability
is only applicable for particles with energy less than
\begin{equation}
\gamma_{max}\simeq1.5\times 10^{-9}[n_{p}^{-1}(V_{ph}/V)(Lv\Omega_0/V^2)^{0.5}]^{1/1.1},
\end{equation}
which for HIM, provides $\sim 20$GeV if taking the injection speed to be $V\simeq 25$km/s. Similar result was obtained with Alfv\'en modes by \citet{FG04}.

Magnetic field itself is likely to be amplified through an inverse 
cascade of magnetic energy at which perturbations created at a particular
$k$ diffuse in $k$ space to smaller $k$ thus inducing inverse cascade. 
As the result, the magnetic perturbations at smaller $k$ get larger than the 
regular field. Consequently, even if the instability is suppressed
for the growth rate given by Eq. (\ref{instability}) it gets efficient
due to the increase of perturbations of magnetic field stemming from the
inverse cascade. The precise picture of the process depends on yet not completely
clear details of the inverse cascade of magnetic field. 

Below, we present the application of the current understanding of the interaction between the streaming instability and the background turbulence to the modeling of the gamma ray emission from molecular clouds near SNRs \citep[see more details in][]{YLS12}. We shall treat the problem in a self-consistent way by comparing the streaming level that is allowed by the preexisting turbulence and the required diffusion for the CRs. 

\subsection{Application to CR acceleration at the shocks }
\label{pmax}
Diffusive shock acceleration of energetic CR particles relies on the crucial process of amplification of MHD turbulence so that particles can be trapped at the shock front long enough to be accelerated to the high energy observed. One of the most popular scenarios that has been adopted in the literature is the streaming instability generated by the accelerated particles. However, in the highly nonlinear regime the fluctuations of magnetic field arising from the streaming
instability get large and the classical treatment of the streaming instability is not applicable. We circumvent the
problem by proposing that the field amplification we consider does not arise from the streaming
instability, but is achieved earlier through other processes, e.g. the interaction of the shock precursor with density perturbations preexisting in the interstellar medium \citep*{BJL09}. Due to the
resonant nature of the streaming instability, the perturbations $\delta B$ arising from it are more efficient
in scattering CRs compared to the large scale fluctuations produced by non-resonant mechanisms, e.g.
the one in \citet{BJL09}. Therefore in this chapter, we limit our discussions to the regime of $\delta B \sim< B_0$, where $B_0$ is the magnetic field that has already been amplified in the precursor region\footnote{The effective $B_0$ is therefore renormalized and can be much larger than the typical field in ISM (see, e.g., \citealt{Diamond_Makov}).}.  

When particles reach the maximum energy at a certain time, they escape and the growth of the streaming instability stops. Therefore we can obtain the maximum energy by considering the stationary state of the evolution. The steady state energy density of the turbulence $W(k)$ at the shock is determined by

\begin{equation}
(U\pm v_A)\nabla W(k) = 2 (\Gamma_{cr}-\Gamma_d)W(k),
\label{wave}
\end{equation}
where $U$ is the shock speed, and the term on the l.h.s. represents the advection of turbulence by the shock flow. $v_A\equiv B_0/\sqrt{4\pi nm}$ and $n$ are the Alfv\'en speed and the ionized gas number density of the precursor region, respectively. The plus sign represents the forward propagating Alfv\'en waves and the minus sign refers to the backward propagating Alfv\'en waves. The terms on the r.h.s. describes the wave amplification by the streaming instability and damping with $\Gamma_d$ as the corresponding damping rate of the wave.  The distribution of accelerated particles at strong shocks is $f(p)\propto p^{-4}$. If taking into account the modification of the shock structure by the accelerated particles, the CR spectrum becomes harder. Assume the distribution of CRs at the shock is $f_0(p)\propto p^{-4+a}$. The nonlinear growth was studied by \citet{Ptuskin:2005}. 

The generalized growth rate of streaming instability is

\begin{eqnarray}
\Gamma_{cr}&=&\frac{12\pi^2 q^2v_A\sqrt{1+A^2}}{c^2k}\nonumber\\
&\times& \int^\infty_{p_{res}} dp p\left[1-\left(\frac{p_{res}}{p}\right)^2\right]D\left|\frac{\partial f}{\partial x}\right|, 
\label{general_growth}
\end{eqnarray}
where $q$ is the charge of the particle, c is the light speed, $p_{res}=ZeB_0\sqrt{1+A^2}/c/k_{res}$ is the momentum of particles that resonate with the waves.  $A=\delta B/B_0$ is wave amplitude normalized by the mean magnetic field strength $B_0$.
\begin{equation}
D=\sqrt{1+A^2}v r_g/3/A^2(>k_{res})
\label{crdiff}
\end{equation}
is the diffusion coefficient of CRs, $v$ and $r_g$ are the velocity and Larmor radius of the CRs.  
In the planar shock approximation, one gets the following growth rate of the upstream forward moving wave at x=0,
\begin{equation}
\Gamma_{cr}(k)=\frac{C_{cr}\xi U^2(U+v_A)k^{1-a}}{(1+A^2)^{(1-a)/2}cv_A\phi(p_{max})r_0^a} 
\label{growth}
\end{equation}
where $C_{cr}=4.5/(4-a)/(2-a)$, $r_0=m c^2/q/B_0$, where $\xi$ measures the ratio of CR pressure at the shock and the upstream  momentum flux entering the shock front, $m$ is the proton rest mass, and $p_{max}$ is the maximum momentum accelerated at the shock front. $H(p)$ is the Heaviside step function.

The linear damping is negligible since the medium should be highly ionized. In fully ionized gas, there is nonlinear Landau damping, which, however, is suppressed due to the reduction of particles' mean free path in the turbulent medium \citep[see][]{YL11}. We therefore neglect this process here. Background turbulence itself can cause nonlinear damping to the waves \citep{YL02}. Unlike hydrodynamical turbulence, MHD turbulence is anisotropic with eddies elongated along the magnetic field. The anisotropy increases with the decrease of the scale \citep{GS95}. Because of the scale disparity, $k_\| > k_\bot \gg k^t_\|$, the nonlinear damping rate in MHD turbulence is less than the wave frequency $k_\| v_A$, and it is given by \citep{FG04, YL04}

\begin{equation}
\Gamma_d \sim \sqrt{k/L} v_A,
\label{damping}
\end{equation}
where L is the injection scale of background turbulence, and the $k$ is set by the resonance condition $k \sim k_\| \sim 1/r_L$.

There are various models for the diffusive shock acceleration. We consider here the escape-limited acceleration. In this model, particles are confined in the region near the shock where turbulence is generated. Once they propagate far upstream at a distance $l$ from the shock front, where the self-generated turbulence by CRs fades away, the particles escape and the acceleration ceases.  The characteristic length that particles penetrate into the upstream is $D(p)/U$. The maximum  momentum is reached when $D(p)/U\simeq l/4$\footnote{The factor 1/4 arises from the following reason. As pointed out by \cite{Ostrowski:1996}, the spectrum is steepened for small l, i.e., $l U/D(p) \sim< 4$}. Assuming $l=\kappa R_{sh}$, where $\kappa<1$ is a numerical factor, one can get
\begin{equation}
\frac{p_{max}}{mc} = \frac{3\kappa A^2 U R_{sh}}{\sqrt{1+A^2}v r_0}.
\label{gmax}
\end{equation}

In particular, for $A<1$
\begin{eqnarray}\frac{p_{max}}{mc}&=&\left[\left(-v_A\sqrt{\frac{1}{r_0 L}}+\sqrt{\frac{v_A^2}{r_0 L}+\frac{2C_{cr}a\xi U^3(U+v_A)}{\kappa r_0R_{sh}cv_A}}\right) \left(\frac{\kappa R_{sh}}{U}\right)\right]^2,\nonumber\\
A&=&\frac{p_{max}r_0}{\sqrt{18}\kappa mU R_{sh}}\sqrt{1+\sqrt{1+36 \left(\frac{\kappa mU R_{sh}}{p_{max}r_0}\right)^2}},
\label{gmax_general}
\end{eqnarray}

\begin{figure}
\includegraphics[width=0.47\textwidth]{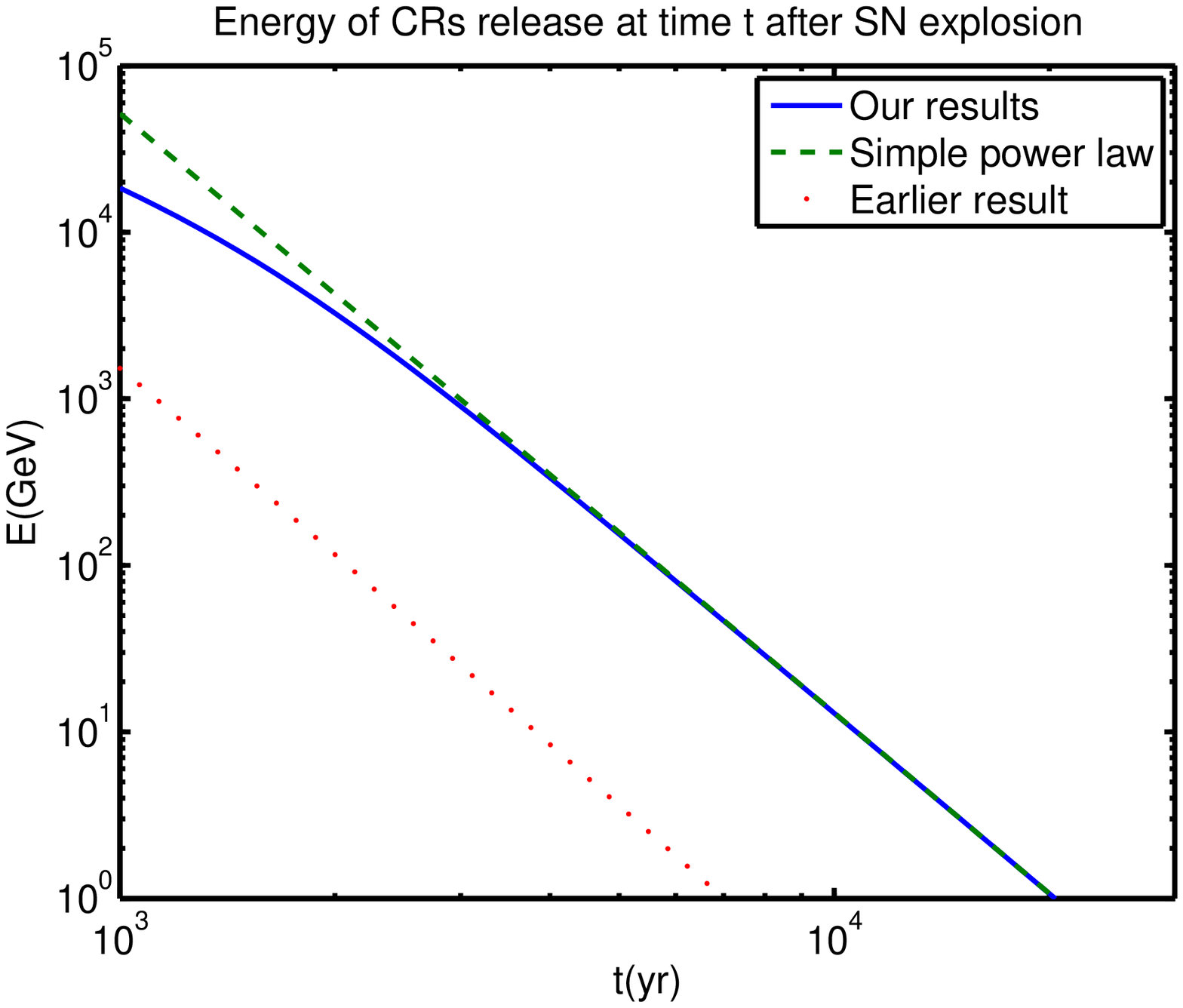}
\includegraphics[width=0.47\textwidth]{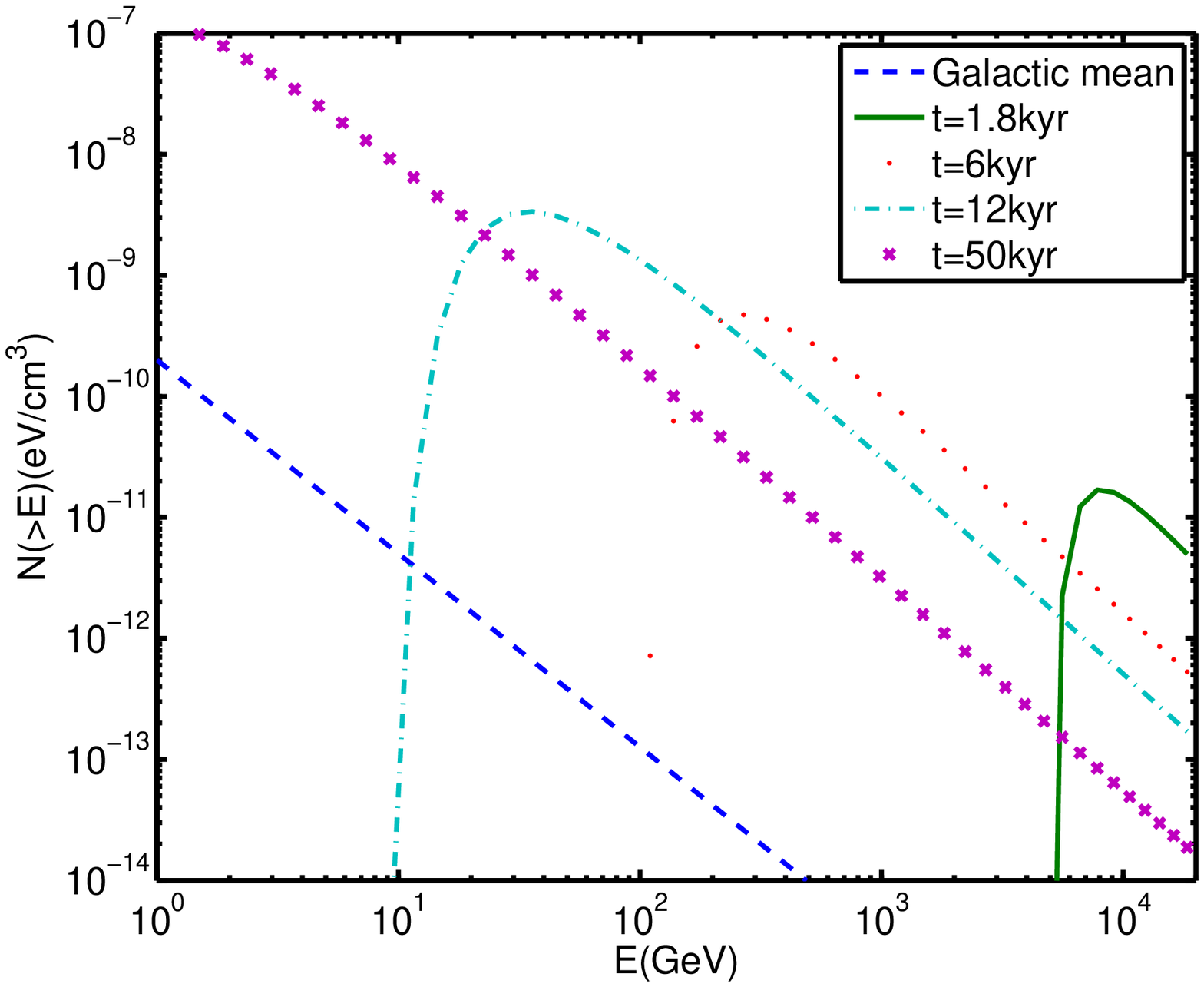}
 \caption{\small {\em Left}: The energy of CRs that are released at the shock at time t in the Sedov phase. Our result shows that the often assumed power law solution \citep[see][]{Gabici:2009, Ohira:2010} is only realized in asymptotic regime as described in Eqs.(\ref{lowshockU},\ref{gmax_solution}). It is also larger than the earlier result (dotted line) in Ptuskin \& Zirakashvili (2005) where the damping of the waves by background turbulence is overestimated. {\em Right}: The spectrum of CRs at a distance $r=12$ pc after 1800 (solid line), 6000 (dotted line), 12000 (dashdot line), 50000 years (cross line). The Galactic mean is plotted as a reference (dashed line). From \cite{YLS12}.}
\label{Emax} 
\end{figure}

In the limit of low shock velocity, 
\begin{eqnarray}
v_A\ll U\ll  c\left[\left(\frac{v_A}{c}\right)^3\frac{\kappa R_{sh}}{2 L C_{cr}a\xi}\right]^{1/4},
\label{lowshockU}
\end{eqnarray}
we get 
\begin{eqnarray}
\frac{p_{max}}{mc}&=&(C_{cr}\xi U^3)^2\frac{a^2 L}{r_0c^2v_A^4}
\label{gmax_solution}
\end{eqnarray}
for the Sedov phase ($t>t_{sed}\equiv 250(E_{51}/(n_0U_9^5))^{1/3}$yr), where $E_{51}=E_{SN}/10^{51}$erg and $U_9=U_i/10^9$cm/s are the total energy of explosion and the initial shock velocity. In Fig.\ref{Emax}, we plot the evolution of $p_{max}/(mc)$ during the Sedov phase. The solid line represents the results from Eqs.(\ref{gmax_general}). As we see, at earlier epoch when advection and streaming instability are both important, the evolution of $p_{max}$ does not follow a power law. For comparison, we also put a power law evolution in the same figure as depicted by Eq.(\ref{gmax_solution})  (dashed line). 
Our result is also larger than that obtained by \cite{Ptuskin:2005} since the wave dissipation rate is overestimated in their treatment.

\subsection{Enhanced scattering and streaming instability near SNRs}
\label{nearby}

The result from \cite{YLS12} show that the local scattering of CRs has to be enhanced by an order of magnitude $\chi =0.05$ in order to produce the amount of $\gamma$ ray emission observed. A natural way to increase the scattering rate is through the streaming instability. The enhanced flux of the CRs are demonstrated to generate strong enough instability to overcome nonlinear damping by the background turbulence \citep{YLS12}. The growth rate in the linear regime is

\begin{equation}
\Gamma_{gr}=\Omega_0\frac{N(\geq E)}{n}\left(\frac{v_s}{v_A}-1\right),
\end{equation}
where $v_s$ is the streaming speed of CRs. The growth rate should overcome the damping rate (eq.\ref{damping}) for the instability to operate. The condition $\Gamma_{gr}>\Gamma_d$ leads to
\begin{equation}
v_s > v_A \left(1+\frac{n v_A}{N \Omega_0\sqrt{r_gL}}\right)
\end{equation}

\begin{figure}
\includegraphics[width=0.47\textwidth]{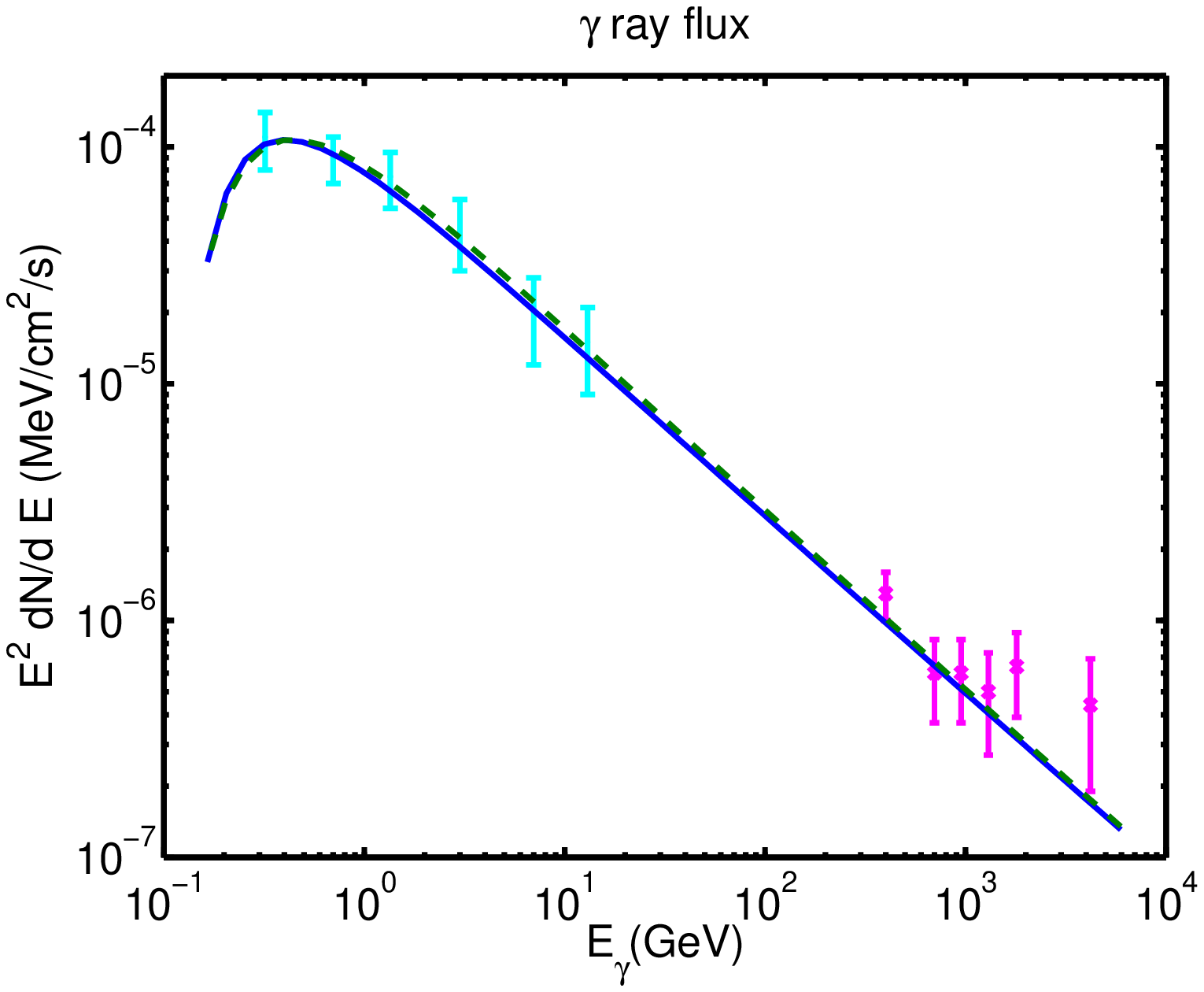}
\includegraphics[width=0.47\textwidth]{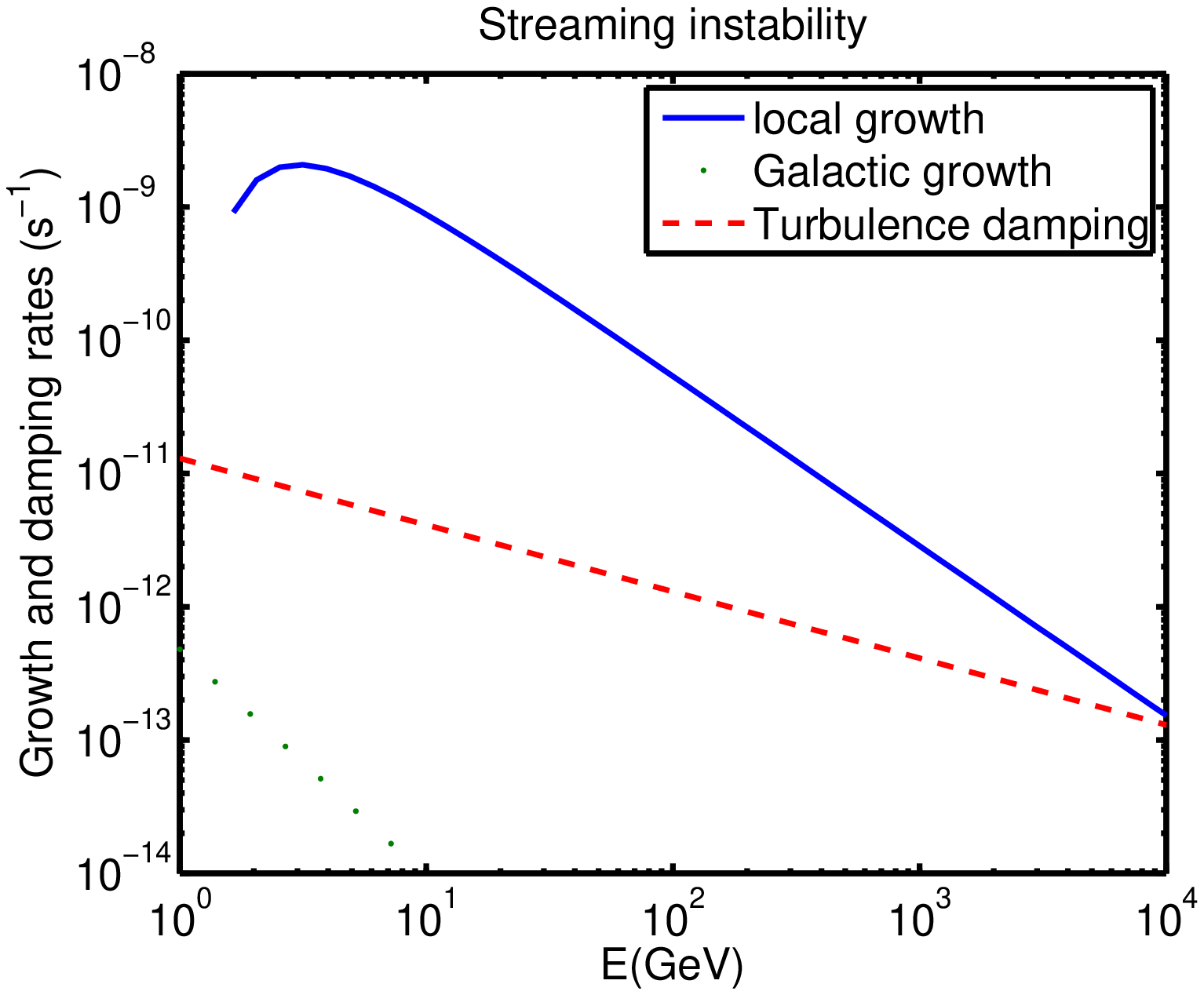}
 \caption{\small {\em Left}: The spectrum of Gamma ray emission from W28. The Fermi data are shown as dotted points \citep{Abdo:2010}, and the H.E.S.S. data are plotted as 'x' points \citep{Aharonian:2008} with error bars. Solid line is our result. {\em Right}: The growth and nonlinear damping rates of streaming instability. With the locally enhanced flux, the growth rate of streaming instability becomes much larger than the mean Galactic value so that it can overcome the nonlinear damping by turbulence for a wide energy range. This is consistent with our earlier treatment in which streaming instability plays an essential role in the cosmic ray diffusion near SNRs. From \cite{YLS12}.}
 \label{rates}
\end{figure}

The spatial diffusion coefficient adopted here, $D \approx v_s L = \chi D_{ISM}$, satisfies  this requirement. The growth and damping rates are compared in Fig.\ref{rates} {\em right}. We see that the streaming instability works in the energy range needed to produce the observed $\gamma$ ray emission, proving that our results are self-consistent.

Note that the case we consider here is different from the general interstellar medium discussed in \cite{YL04} and \cite{FG04}, namely, the local cosmic ray flux near SNRs is much enhanced (see Fig.\ref{Emax} {\em right}). Consequently, the growth rate of the streaming instability becomes high enough to overcome the damping rate by the preexisting turbulence in the considered 
energy range.

\begin{table}
\caption{Model parameters adopted}
\begin{tabular}{cccccc}
\hline
\hline
a&$\chi$&$\eta$&$ \kappa$&$\xi$&$\alpha$\\
\hline
0.1$\sim 0.3$&$\sim$0.05&$\sim 0.3$&$0.04\sim 0.1$& $0.2\sim$0.4& 0.5\\
\hline
\hline
\end{tabular}
\end{table}

\section{Gyroresonance Instability of CRs in Compressible Turbulence}

\begin{figure}
\includegraphics[width=0.55\columnwidth,
  height=0.25\textheight]{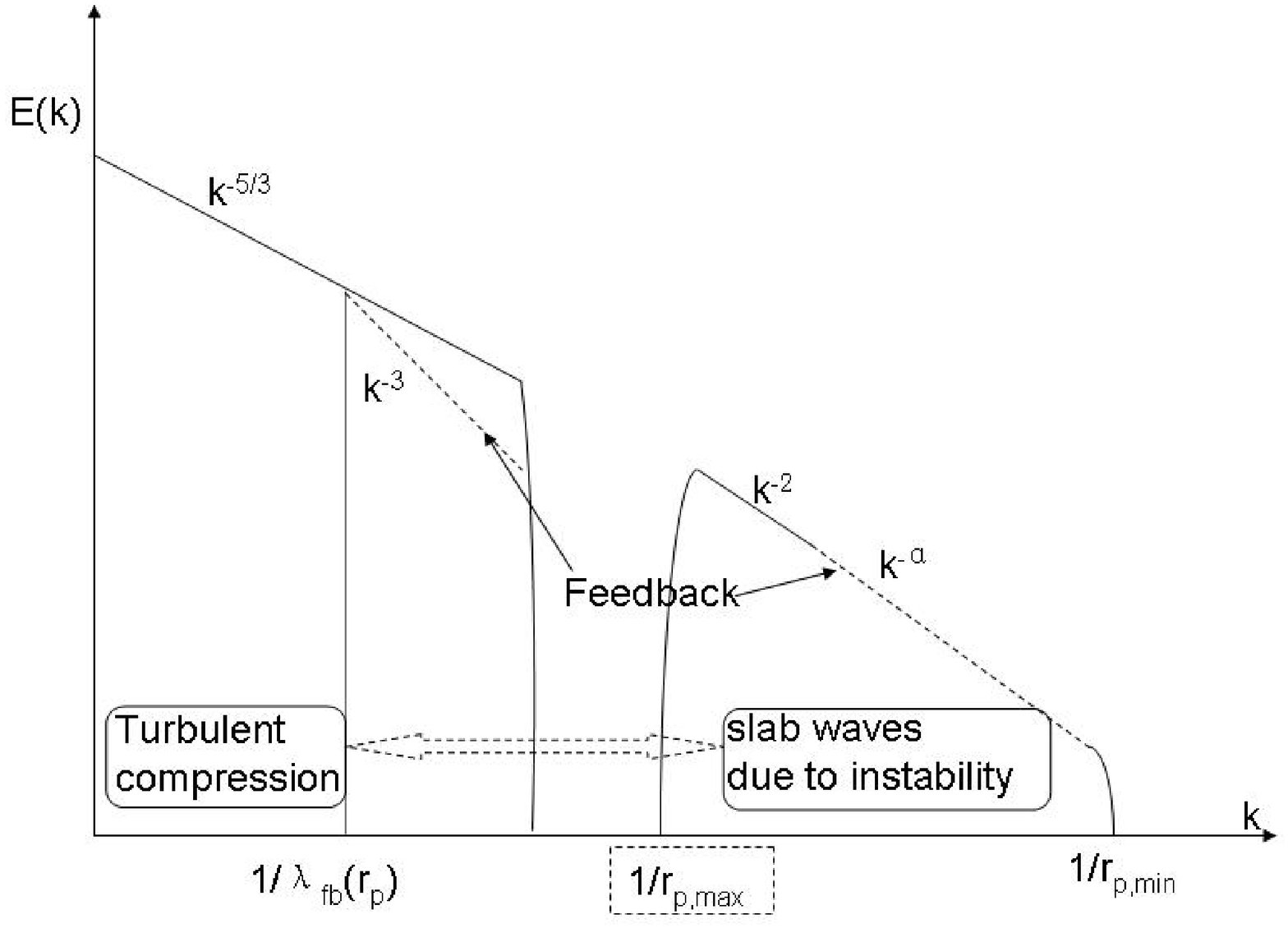}
\includegraphics[width=0.4\columnwidth,
  height=0.25\textheight]{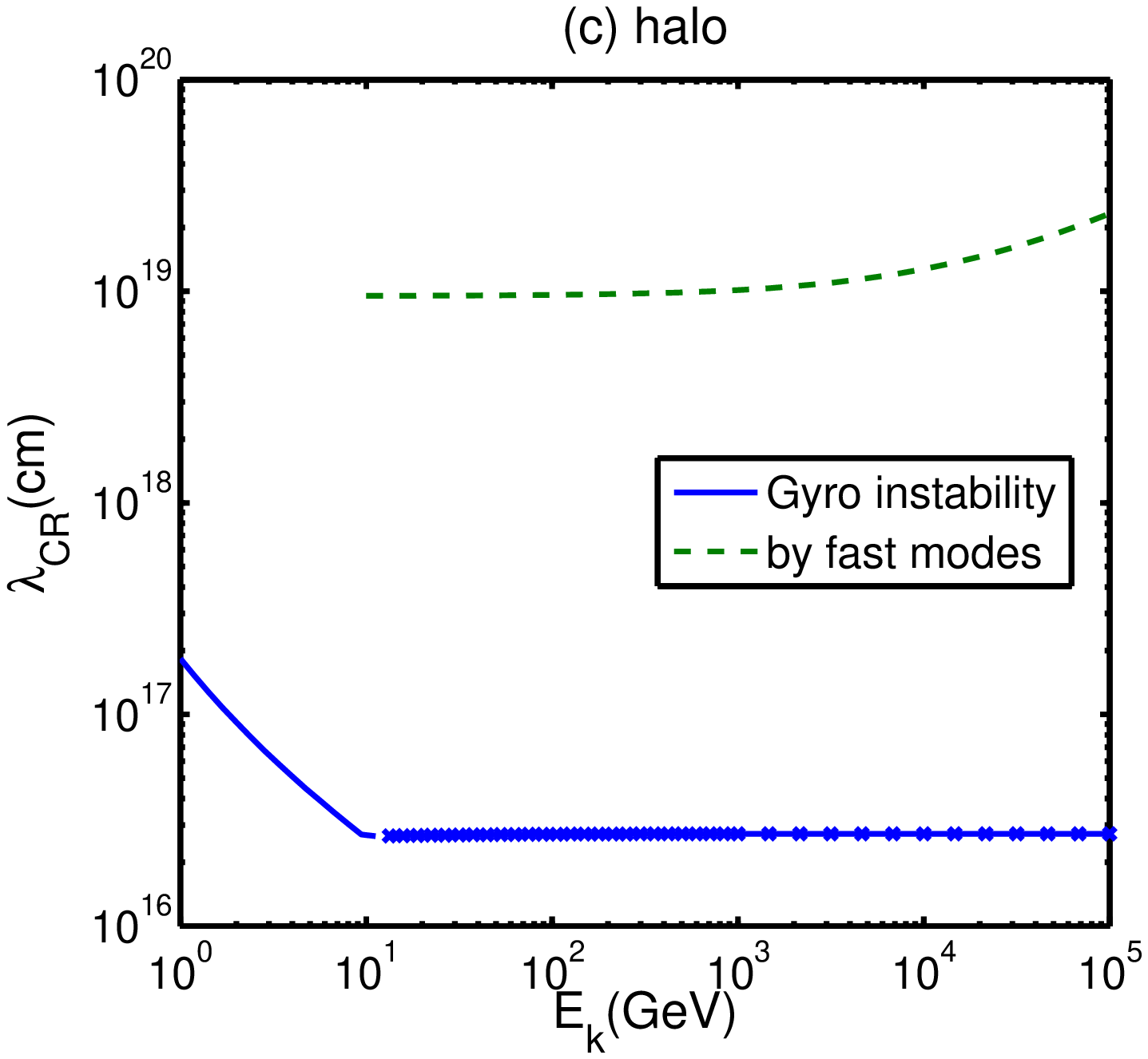}
\caption{{\em Left}: The spectral energy density of slab waves that is transferred from the large scale compressible turbulence via the gyroresonance instability of CRs. In the case that the instability grows up to the maximum energy rate allowed by the turbulence cascade, large scale turbulence is truncated at $\lambda_{fb}$ and the wave amplitude $E(k)dk\sim \epsilon_N^u$ is given by Eq.(\ref{energy}). Note that the picture is different from LB06, namely, the feedback on the large scale turbulence occurs only in some cases when the scattering is not sufficient to prevent the waves from growing to the maximum values \citep{YL11}; {\em right}: CR scattering is dominated by compressible modes. For high energy CRs ($>\sim$10GeV), the scattering is due to direct interaction with fast modes; For low energy CRs, the interaction is mainly due to the gyroresonance instability induced by compression of magnetic fields.}
\label{feedback_fig}
\end{figure}

Until recently, test particle approximation was assumed in most of earlier studies in which turbulence cascade is established from large scales and no feedback of CRs is included. This may not reflect the reality as we know the energy of CRs is comparable to that in turbulence and magnetic field \citep[see][]{Kulsrudbook}. It was suggested by \cite{LB06} that the gyroresonance instability of CRs can drain energy from the large scale turbulence and cause instability on small scales by the turbulence compression induced anisotropy on CRs (see Fig.\ref{feedback_fig} {\em left}). And the wave generated on the scales, in turn, provides additional scattering to CRs. In \cite{YL11}, quantitative studies was provided based on the nonlinear theory of the growth of the instability and the feedback of the growing waves on the distributions of CRs.

In the presence of background compressible turbulence, the CR distribution is bound to be anisotropic because of the conservation of the first adiabatic invariant $\mu\equiv v_\bot^2/B$. Such anisotropic distribution is subjected to various instabilities. Waves are generated through the instabilities, enhancing the scattering rates of the particles, their distribution will be relaxed to the state of marginal state of instability even in the collisionless environment.  While the hydrodynamic instability requires certain threshold, the kinetic instability can grow very fast with small deviations from isotropy. Here, we focus on the gyroresonance instability. Both the qualitative and quantitative studies in \cite{YL11} show that the isotropization rate is roughly $
\tau^{-1}_{scatt} \sim \frac{\Gamma_{gr}\epsilon_N}{\beta_{CR} A}\label{nu_est}$, where $\Gamma_{cr}, \epsilon_N$ are the instability growth rate and the wave energy normalized by magnetic energy, respectively. $\beta_{CR}$ is the ratio of CR pressure to magnetic pressure, $A$ is the degree of anisotropy of the CR momentum distribution.

By balancing the rate of decrease in anisotropy due to scattering and the growth due to compression, one can get
\be
\epsilon_N\sim \frac{ \beta_{CR}\omega\delta v}{\Gamma_{gr} v_A },~~~\lambda_{CR}=r_p/\epsilon_N.
\label{epsilon_est},
\ee
where $v_A$ is the Alfv\'en speed, $\omega, \delta v$ are the wave frequency and amplitude at the scale that effectively compresses the magnetic field and create anisotropy in CRs' distribution \citep{YL11}. Since the growth rate decreases with energy, the instability only operates for low energy CRs ($\sim<$ 100GeV, see Fig.\ref{feedback_fig} {\em right}) due to the damping by the preexisting turbulence \citep{YL11} .
  
\subsection{Bottle-neck for the growth of the instability and feedback on turbulence}
\label{feedback}
The creation of the slab waves through the CR resonant instability is another channel to drain the energy of large scale turbulence. This process, on one hand, can damp the turbulence. On the other hand, it means that the growth rate is limited by the turbulence cascade. The energy growth rate  cannot be larger than the turbulence energy cascading rate, which is $1/2 \rho V_L^4/v_A/L$ for fast modes in low $\beta$ medium and $\rho v_A^3/l_A$ for slow modes in high $\beta$ medium. This places a constraint on the growth, thus the upper limit of wave energy is given by
\bea
\epsilon^u_N=\cases{ M_A^2 L_i/(L A)\gamma^{\alpha-1},& $\beta<1$ \cr
  L_i/(l_A A)\gamma^{\alpha-1}, & $\beta>1$, \cr}
\label{energy}
\eea
where $\gamma$ is the Lorentz factor and $L_i\simeq 6.4\times 10^{-7}(B/5{\rm \mu G})(10^{-10}{\rm cm}^3/n_{cr})$ pc. The growth is induced by the compression at scales $\sim< \lambda_{CR}$. Therefore, in the case that $\Gamma_{gr} \epsilon$ reaches the energy cascading rate, fast modes are damped at the corresponding maximum turbulence pumping scale $\lambda_{fb}=r_p/\epsilon_N$ (see Fig.\ref{feedback_fig} {\em left}).  If $\lambda_{fb}$ is larger than the original damping scale $l_c$, then there is a feedback on the large scale compressible turbulence. This shows that test particle approach is not adequate and feedback should be included in future simulations.


\section{Summary}

In this chapter, we reviewed recent development on cosmic ray transport theories based on modern understanding of MHD turbulence. The main conclusions from both analytical study and test particle simulations in MHD turbulence are:
\begin{itemize}
\item Compressible fast modes are most important for CR scattering. CR transport therefore varies from place to place.
\item Nonlinear mirror interaction is essential for pitch angle scattering (including 90 degree).
\item Cross field transport is diffusive on large scales and super-diffusive on small scales.
\item Subdiffusion does not happen in 3D turbulence.
\item Self - generated waves are subject to damping by preexisting turbulence 
\item Small scale waves can be generated in compressible turbulence by gyroresonance instability. Feedback of CRs on turbulence need to be included in future simulations. 
\end{itemize}

\bibliographystyle{apj}
\bibliography{yan}


\end{document}